\documentclass[aps,prb,twocolumn,showpacs,amsfonts,%
superscriptaddress]{revtex4-1}

\usepackage{amsmath}
\usepackage{graphicx}
\usepackage{color}
\usepackage{hyperref}
\usepackage{sidecap}
\usepackage{mathbbol}

\newcommand{\beq}{\begin{equation}}
\newcommand{\eeq}{\end{equation}}
\newcommand{\beqarray}{\begin{eqnarray}}
\newcommand{\eeqarray}{\end{eqnarray}}
\newcommand{\mbs}[3] {_{#1,{\bf #2},#3}}
\newcommand{\mom}[1] {_{{\bf #1}}}
\newcommand{\bamo}[2] {_{#1,{\bf #2}}}
\newcommand{\trafo}[6]{{\cal U}^{#1}_{#2, #3; #4}({\bf #5}, #6)}
\newcommand{\prevl}[3]{Phys.\ Rev.\ Lett.\ {\bf #1}, #2 (#3).}
\newcommand{\prevb}[3]{Phys.\ Rev.\ B {\bf #1}, #2 (#3).}

\allowdisplaybreaks

\begin{document}

\title{Superconducting pairing in the spin-density-wave phase of iron
pnictides}

\author{Jacob Schmiedt}
\email{jacob\_alexander.schmiedt@tu-dresden.de}
\affiliation{Institute of Theoretical Physics, Technische Universit\"{a}t
Dresden, 01062 Dresden, Germany}
\author{P. M. R. Brydon}
\affiliation{Condensed Matter Theory Center and Joint Quantum Institute,
Department of Physics,
University of Maryland, College Park, Maryland 20742-4111, USA}
\affiliation{Institute of Theoretical Physics, Technische Universit\"{a}t
Dresden, 01062 Dresden, Germany}
\author{Carsten Timm}
\email{carsten.timm@tu-dresden.de}
\affiliation{Institute of Theoretical Physics, Technische Universit\"{a}t
Dresden, 01062 Dresden, Germany}

\date{\today}

\begin{abstract}
Some of the iron pnictides show coexisting superconductivity and
spin-density-wave order. We study the superconducting pairing
instability in the spin-density-wave phase. Assuming that the pairing
interaction is due to spin fluctuations, we calculate the effective
pairing interactions in the singlet and triplet channels by summing
the bubble and ladder diagrams taking the reconstructed band structure
into account. The leading pairing instabilities and the corresponding
superconducting gap structures are then obtained from the
superconducting gap equation. We illustrate this approach for a
minimal two-band model of the pnictides. Analytical and numerical
results show that the presence of spin and charge fluctuations in the
spin-density-wave phase strongly enhances the pairing. Over a limited
parameter range, a $p_x$-wave state is the
dominant instability. It competes with various states, which have 
mostly $s^\pm$-type structures. We analyze the effect of
various symmetry-allowed interactions on the pairing in some detail.
\end{abstract}


\maketitle

\section{Introduction}\label{intro}

Understanding the phase diagrams of iron-pnictide superconductors has been
an important challenge for the condensed matter community in
recent years.\cite{doi:10.1080/00018732.2010.513480, 0034-4885-74-12-124508,
0953-8984-22-20-203203} This large class of compounds can be subdivided into
several families according to their crystal structure.
Among the most intensively studied families are the so-called 1111 compounds
$R\mathrm{FeAsO}$, where $R$ is a rare-earth element,
and the 122 compounds $A\mathrm{Fe_2As_2}$, where $A$ is an
alkaline-earth element. The materials in these families share
important features: The undoped parent compounds show
an antiferromagnetic spin-density-wave (SDW) phase below a N\'eel
temperature $T_N$ and a structural transition from a tetragonal to an
orthorhombic phase at the same or a slightly higher temperature. The magnetic
and orthorhombic phase is suppressed by doping or by applying
pressure. Close to where the magnetic and structural phase
transitions approach zero temperature, superconductivity
appears.\cite{0953-8984-22-20-203203, Luetkens2009, PhysRevLett.103.087001,
PhysRevLett.107.237001, PhysRevB.83.172503, PhysRevB.80.024508,
PhysRevLett.101.057006} The proximity of the magnetic and superconducting (SC)
phases suggests a close relationship between the two phenomena. Hence, spin
fluctuations are widely considered to provide the pairing ``glue''
in these systems,\cite{1367-2630-11-2-025016, PhysRevB.79.224511,
PhysRevB.81.214503, PhysRevB.81.054502} although it has also been proposed that
orbital fluctuations are critical for the
superconductivity.\cite{PhysRevB.88.045115, PhysRevLett.104.157001,
PhysRevB.82.144510}
It has been shown that the
spin-fluctuation-mediated pairing interaction in the paramagnetic phase of the
iron pnictides is
repulsive between electron and hole Fermi pockets.\cite{1367-2630-11-2-025016,
PhysRevB.79.224511, PhysRevB.81.214503, PhysRevB.81.054502, PhysRevB.82.024508}
Therefore, sign changes in the gap are required to satisfy the BCS gap equation,
which leads to an $s^\pm$-state as the dominant SC instability.

The underdoped region of the phase diagram, close to the disappearance of
the SDW, is particularly interesting. Intuitively, one might expect that the SDW
and superconductivity should not coexist because both types of
order compete for the same electrons. Indeed, for
fluorine-doped $\mathrm{LaFeAsO}$ under ambient pressure, a strong first-order
transition between the SDW phase and the SC phase is
observed.\cite{Luetkens2009} In
other cases, e.g., for fluorine-doped $\mathrm{LaFeAsO}$ under
pressure\cite{PhysRevB.84.100501} and for
$\mathrm{CaFe_2As_2}$,\cite{PhysRevB.80.024508}
the two phases coexist but are thought to be separated into different domains on
a mesoscopic scale. On the other hand, for
hole-doped $\mathrm{Ba}_{1-x}\mathrm{K}_x\mathrm{Fe_2As_2}$ and electron-doped
$\mathrm{Ba}(\mathrm{Fe}_{1-x}\mathrm{Co}_x)_2\mathrm{As}_2$ there is strong
experimental evidence from X-ray diffraction,\cite{PhysRevLett.103.087001}
neutron
scattering,\cite{PhysRevLett.103.087001, PhysRevB.81.140501, PhysRevB.83.172503}
NMR,\cite{0295-5075-87-3-37001, PhysRevB.86.180501} and 
$\mu$SR\cite{PhysRevLett.107.237001} that
the SDW, superconductivity, and the orthorhombic distortion coexist
microscopically.
In these systems, there exists a finite doping range where upon cooling the
system first undergoes the structural and magnetic transitions and at a lower
temperature becomes superconducting. The SDW order displays reentrant behaviour 
in the $\mathrm{Ba}(\mathrm{Fe}_{1-x}\mathrm{Co}_x)_2\mathrm{As}_2$ system,
disappearing at still lower temperature.\cite{PhysRevB.81.140501}

Studies based on microscopically derived Ginzburg-Landau
functionals\cite{PhysRevB.82.014521} find that due to the
multiband nature of the pnictides a conventional \textit{s}-wave SC state
with the same sign of the SC gap on all Fermi pockets and the SDW are mutually 
exclusive. On the other hand, a $s^\pm$-state with opposite signs of the gap on
electron and hole Fermi pockets can coexist with a SDW. These results are
consistent with 
mean-field calculations inside the coexistence phase which find coexistence of
the SDW with a $s^\pm$-state to be much more favorable than with a conventional
$s$-wave state.\cite{PhysRevB.80.100508, PhysRevB.83.224513, PhysRevB.81.174538}
It has also been shown that an
increasing magnitude of the SDW amplitude can lead to the appearance of
accidental nodes of the SC gap in the coexistence
regime,\cite{PhysRevB.85.144527} which could explain
thermal-conductivity measurements suggesting vertical line
nodes in strongly underdoped
$\mathrm{Ba}_{1-x}\mathrm{K}_x\mathrm{Fe_2As_2}$.\cite{arXiv:1105.2232} 
However, these theoretical works either assume a simple
phenomenological pairing interaction\cite{PhysRevB.82.014521} or consider only
the bare electron-electron interaction\cite{PhysRevB.81.174538,
PhysRevB.85.144527} to obtain pairing in the SDW phase.
They do not consider any momentum dependence of the interaction beyond the one
resulting from the unitary transformation onto reconstructed bands in the SDW
phase. Although it is generally recognized that
spin fluctuations are crucial for the understanding of superconductivity in the
paramagnetic phase, their effect in the SDW phase has not received a lot of
attention. In particular, the breaking of spin-rotation symmetry leads to the
appearance of propagating magnon modes and the presence of these modes is
expected to strongly affect the pairing. It is not covered
by the approaches discussed above. In a first attempt to include magnetic
excitations in the SDW phase, Wu and Phillips\cite{0953-8984-23-9-094203} have
studied a spin-fermion model with a single electronic band
coupled to localized spins. This
approach also gives an $s^\pm$-state as the leading SC instability in the SDW
phase. However, it does not take into account that the same particles are
responsible for the formation of collective magnetic excitations and of Cooper
pairs. A more realistic description should be based on the pairing interaction
due to the exchange of spin fluctuations, calculated for a multiband
electronic model in the magnetically ordered state. Our goal is to develop such
a description.

Our approach consists of two steps: First, we obtain the approximate effective
pairing interaction in the presence of the SDW by summing up the bubble and
ladder diagrams in the
particle-hole channel that contribute to the effective pairing vertex. This
allows
us to express the pairing interaction in terms of
random-phase-approximation (RPA) susceptibilities. Second, we follow Berk
and Schrieffer\cite{PhysRevLett.17.433} by inserting the
pairing interaction into the linearized BCS gap equation to obtain the leading
SC instability. This approach has been used extensively to study
SC pairing in the paramagnetic phase of the iron
pnictides.\cite{1367-2630-11-2-025016, PhysRevB.79.224511, PhysRevB.81.214503}

In the first part of the paper, Sec.\ \ref{method}, we develop this approach for
a multiband system in the presence of a SDW. We thereby fill the gap between
earlier works that either obtain the effective pairing interaction in the
presence of
SDW order for a one-band model\cite{PhysRevB.39.11663, PhysRevB.46.11884}
or that apply the RPA to multiband systems in the absence of
long-range order.\cite{1367-2630-11-2-025016, PhysRevB.79.224511,
PhysRevB.81.214503}
We note already here that an important consequence of the breaking of
spin-rotation symmetry by the SDW is the mixing of spin-singlet and spin-triplet
pairing. Therefore, the naive spin degree of freedom of the
quasiparticles in the SDW phase is not the same as the bare electron spin.
We will call the former the ``quasi-spin.'' The definition will be made more
precise below.

In the rest of the paper, we apply this technique to a two-band model
with momentum-independent interactions, which is introduced in Sec.\
\ref{model}. Our model is inspired by two-band models that are frequently used
as minimal models for the iron
pnictides\cite{PhysRevB.78.134512, PhysRevB.80.174401, PhysRevB.84.180510,
PhysRevB.84.214510} because they reproduce central features of the Fermi
surface: They include one hole Fermi pocket around $(0,0)$ and
two electron Fermi pockets around $(\pi,0)$ and $(0,\pi)$ in the unfolded
Brillouin zone (BZ). We then study the effect of various symmetry-allowed
types of bare interactions on the effective pairing interaction and the
resulting SC gap structure, using analytical arguments in Sec.\
\ref{analytical} and numerical calculations in Sec.\ \ref{numerics}. We pay
particular attention to the effect of the magnons in the SDW phase since
they lead to a divergence of the interband components of the transverse RPA spin
susceptibility. As predicted by previous
studies,\cite{PhysRevB.82.014521, PhysRevB.81.174538} the dominant
quasi-spin-singlet state has an $s^\pm$-type structure. However, we find
extended parameter ranges where a quasi-spin-triplet $p_x$-wave state is the
dominant SC instability. We observe that an interband pair-hopping interaction
is crucial for stabilizing quasi-spin-singlet pairing. In Sec.\ \ref{summary},
we summarize our results and draw some conclusions.

\section{Method}\label{method}

\subsection{Multi-band model with SDW order}\label{saddle-pt}

We introduce our method for a general Hubbard-type model with $N$ bands in the
paramagnetic phase. For simplicity, the interactions are assumed to be momentum
independent in the basis that diagonalizes the free Hamiltonian but are
otherwise general. The
generalization to momentum-dependent interactions, which may arise due to
orbital degrees of freedom, is straightforward. We set $\hbar=1$
and, in the present section, employ the functional-integral formalism.
The action for our model reads
\begin{eqnarray}
S &=& \int_0^\beta \! d\tau\, \Bigg[
  \sum\mom{k,\sigma} \sum_{A} c^*\mbs{A}{k}{\sigma}\, (\partial_\tau +
  \epsilon_{A,{\bf k},\sigma})\, c\mbs{A}{k}{\sigma}\nonumber\\
&& {}+ \frac{1}{2}\sum\mom{k,k',q} \sum_{A,B,C,D} \sum_{\sigma,\sigma'}
  U_{(A,B),(C,D)}(\sigma,\sigma')\nonumber\\
&& {}\times c^*\mbs{A}{k+q}{\sigma}c^*\mbs{C}{k'-q}{\sigma'}
  c\mbs{D}{k'}{\sigma'}c\mbs{B}{k}{\sigma}\Bigg]
  \equiv S_0 + S_\mathrm{int} ,\quad
\label{Sint}
\end{eqnarray}
where the capital letters \textit{A}, \textit{B}, \textit{C},
\textit{D} label the bands in the paramagnetic state and
$c_{A,\mathbf{k},\sigma}$ etc.\ are Grassmann variables.

In the SDW phase above the SC transition temperature, the SDW is the only
electronic order present. The interaction term can be written as
\begin{equation}
S_{\rm int} = S_{\rm SDW} + \Delta S,
\end{equation}
where $S_{\rm SDW}$ is the interaction in the spin channel, which leads to the
formation of the SDW, and $\Delta S$ contains all remaining interaction terms.
The spin-density interaction reads
\begin{equation}
S_{\rm SDW} = \int_0^\beta \! d\tau\, \sum\mom{q}\sum_{A,B,C,D}
  {\bf S}_{AB,-{\bf q}}\, \hat{U}_{ABCD}^{\rm spin}\, {\bf S}_{CD,{\bf q}} ,
\end{equation}
where
\begin{eqnarray}
{\bf S}_{AB,{\bf q}} &=& \frac{1}{2}\sum\mom{k}\sum_{\sigma,\sigma'}
  \left[c^*\mbs{A}{k}{\sigma}\,\frac{\pmb{\sigma}_{\sigma\sigma'}}{2}\,
  c\mbs{B}{k-q}{\sigma'}\right.\nonumber\\
&& \left. {}+ c^*\mbs{B}{k}{\sigma}\,\frac{\pmb{\sigma}_{\sigma\sigma'}}{2}\,
  c\mbs{A}{k-q}{\sigma'}\right]
\end{eqnarray}
and $\pmb{\sigma}$ is the vector of Pauli matrices. $\hat{U}_{ABCD}^{\rm spin}$
are matrices of coupling constants, which can
be obtained from the coefficients $U_{(A,B),(C,D)}(\sigma,\sigma')$ in
Eq.\ (\ref{Sint}). If the interactions do not break spin-rotation invariance
we can write
\begin{equation}
\hat{U}^{\mathrm{spin}}_{(A,B),(C,D)} =
  U_{(A,D),(C,B)}(\uparrow,\downarrow)\,\hat{1}_3,
\end{equation}
where $\hat{1}_3$ is the three-dimensional unit matrix. The
interaction $S_{\rm SDW}$ is decoupled by the introduction of
Hubbard-Stratonovic fields $\pmb{\Delta}_{AB,{\bf q}}$. We assume a
finite static saddle-point value $\pmb{\Delta}_{AB,{\bf Q}} =
\Delta_{AB}\hat{\bf e}_z$
only for $\mathbf{q}=\mathbf{Q}$, with the SDW ordering vector ${\bf Q}$.
The saddle-point SDW order parameters $\Delta_{AB}$ are
obtained from the stationarity conditions of the resulting free energy
${\partial F_\mathrm{sp}}/{\partial \Delta_{AB}}
  \equiv - \beta^{-1}\, {\partial \ln
  Z_\mathrm{sp}}/{\partial \Delta_{AB}} = 0$,
where $Z_\mathrm{sp}$ is the partition function evaluated at the
saddle-point values $\Delta_{AB}$ of the Hubbard-Stratonovic fields.

Fluctuations of the decoupling field around this saddle point are denoted by
$\pmb{\delta}_{AB,{\bf q}}$ so that $\pmb{\Delta}_{AB,{\bf q}} =
\Delta_{AB}\hat{\bf e}_z\delta_{\bf q,Q}+\pmb{\delta}_{AB,{\bf q}}$.
Sufficiently deep in the SDW phase, we can neglect the fluctuations
$\pmb{\delta}\bamo{AB}{Q}$ in the $\mathbf{q}=\mathbf{Q}$ channel compared to
the saddle-point value $\Delta_{AB}\hat{\bf e}_z$---this constitutes the
mean-field approximation for the order parameter. However, we keep the
fluctuations in all other channels, where the saddle-point value is zero.
With this, the action becomes
\begin{eqnarray}
S' &=& S_0 + \Delta S  \nonumber\\
&& {}+ \sum_{A,B,C,D}\Bigg\{\int_0^\beta \! d\tau \sum\mom{q\neq Q}\Big[
  2\pmb{\delta}_{AB,{\bf q}}\cdot {\bf S}\bamo{CD}{-q} \nonumber\\
&& {}- \pmb{\delta}\bamo{AB}{-q}\, (\hat{U}^{\rm
  spin}_{ABCD})^{-1}\, \pmb{\delta}\bamo{CD}{q}\Big]
+ 2\Delta_{AB}\,\hat{\bf e}_z\cdot {\bf S}\bamo{CD}{Q}\nonumber\\
  && {}- \Delta_{AB}\big[
  (\hat{U}^{\rm spin}_{ABCD})^{-1}\big]_{zz}\Delta_{CD}\!\Bigg\}.
\end{eqnarray}
The fluctuation fields can now be integrated out again. This gives the action in
terms of the fermionic fields in the presence of a SDW as
\begin{eqnarray}
S'' &=& S_0 + \sum_{A,B,C,D}\Bigg\{\int_0^\beta \! d\tau \Big[ 2\Delta_{AB}\,
  \hat{\bf e}_z \cdot {\bf S}\bamo{CD}{Q}\nonumber\\
&& {}+ \sum\mom{q\neq Q} \mathbf{S}\bamo{AB}{-q}\, \hat{U}^{\rm
  spin}_{ABCD}\, \mathbf{S}\bamo{CD}{q} \Big]\nonumber\\
&& {}- \Delta_{AB}\big[(\hat{U}^{\rm
  spin}_{ABCD})^{-1}\big]_{zz}\Delta_{CD}\!\Bigg\} + \Delta S.
\end{eqnarray}
In the thermodynamic limit, the sum over $\mathbf{q}$ is replaced by an
integral, for which the exclusion of the single point $\mathbf{q}=\mathbf{Q}$
does not make a difference, unless the integrand is too strongly divergent at
this point. We will show in Secs.\ \ref{analytical} and \ref{numerics} that the
effective interactions remain finite as this point is approached. Hence, we can
drop the exclusion of $\mathbf{q}=\mathbf{Q}$ without changing the results.

The bilinear part of the action $S''$, which consists of $S_0$ and a
contribution from the saddle point, can be diagonalized by a unitary
transformation,
\begin{equation}
c_{A,{\bf k}+n{\bf Q},\sigma} = \sum_{\nu=1}^{2N}\, {\cal U}_{A,n;\nu}({\bf
  k},\sigma)\, d\mbs{\nu}{k}{\sigma},
\label{UTrafo}
\end{equation}
where $n$ can be 0 or 1 and $\nu$ labels the reconstructed bands. We have here
assumed that the SDW doubles the size of the unit cell in real space and thus
halves the BZ and doubles the number of bands. 
Note that the transformation factors depend on the spin index $\sigma$.
Therefore, part of the spin information in the original basis is transferred to
the band information in the new basis. The spin index of the transformed
operator $d\mbs{\nu}{k}{\sigma}$ thus does not contain the full spin
information. We therefore call the quantity $\sum_{\sigma,\sigma'}
d^*\mbs{A}{k}{\sigma}\,(\pmb{\sigma}_{\sigma\sigma'}/2)\,
d\mbs{B}{k-q}{\sigma'}$ a ``quasi-spin.''

Combining the interactions in the SDW channel and in $\Delta S$
into one term again, the action in the new basis becomes
\begin{widetext}
\begin{eqnarray}
S'' &=&  \int_0^\beta \! d\tau\, \Bigg[ \sum_{\nu}
  {\sum_\mathbf{k}}' \sum_\sigma d^*\mbs{\nu}{k}{\sigma}\, (\partial_\tau +
  E_{\nu,{\bf k}})\, d\mbs{\nu}{k}{\sigma} + \frac{1}{2}
  {\sum_{\mathbf{k},\mathbf{k}',\mathbf{q}}}^{\!\!\prime}\,
  \sum_{\sigma,\sigma'}\sum_{j,k,l}\sum_{\nu,\mu,\alpha,\beta}
  \sum_{A,B,C,D} U_{(A,B),(C,D)}(\sigma,\sigma') \nonumber \\
&& {}\times \trafo{*}{A}{|j-l|}{\nu}{k+q}{\sigma}\,
  \trafo{}{B}{j}{\mu}{k}{\sigma}\,
  \trafo{*}{C}{|k-l|}{\alpha}{k'-q}{\sigma'}\,
  \trafo{}{D}{k}{\beta}{k'}{\sigma'}\,
  d\mbs{\nu}{k+q}{\sigma}^* d\mbs{\alpha}{k'-q}{\sigma'}^*
  d\mbs{\beta}{k'}{\sigma'} d\mbs{\mu}{k}{\sigma}\Bigg] \nonumber \\
&& {}- \Delta_{AB}\left[(\hat{U}^{\rm
  spin}_{ABCD})^{-1}\right]_{zz}\Delta_{CD},
\label{S_final}
\end{eqnarray}
\end{widetext}
where $E_{\nu,{\bf k}}$ is the dispersion of the reconstructed bands and
$\sum\mom{k}'$ denotes the sum over the magnetic BZ.

\subsection{Effective pairing interaction and gap equation}

\begin{figure*}
\includegraphics[clip,width=\textwidth]{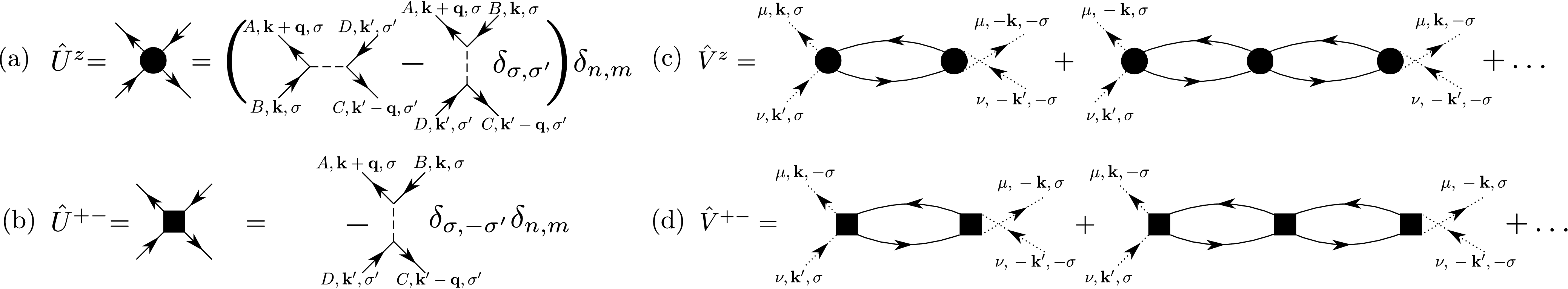}
\caption{Diagrammatic representations of the matrix elements of (a) $\hat{U}^z$
and (b) $\hat{U}^{+-}$. The two lowest-order contributions to the
RPA series for the effective pairing interactions for pairs with vanishing total
quasi-spin $s_z=0$ are shown in (c) for longitudinal
particle-hole fluctuations and in (d) for transverse fluctuations. The dotted
lines represent the transformation factors $\mathcal{U}$ attached to
the external legs.}
\label{RPA}
\end{figure*}

In this section, we calculate the effective pairing
interaction $\Gamma_{\nu,\mu}({\bf k,k'})$ in the presence of the SDW but above
the SC critical temperature $T_c$. The pairing interaction
is then inserted into the linearized gap
equation,\cite{PhysRevLett.17.433} which can be expressed as an eigenvalue
problem with the pairing-symmetry functions $\gamma_\alpha({\bf k})$ as
eigenvectors:
\begin{equation}
-\sum_j\oint_{C_j} \frac{dk'_\parallel}{2\pi v_F({\bf k'})}\;
  \Gamma_{\nu_i,\mu_j}({\bf k,k'})\, \gamma_\alpha({\bf k'}) =
  \lambda_\alpha\, \gamma_\alpha({\bf k}).
\label{gapeq}
\end{equation}
Herein, $v_F({\bf k}) = |\nabla_{\bf k} E_\nu({\bf k})|$ is the Fermi velocity.
The indices $i$ and $j$ label the Fermi pockets and $\nu_i$ denotes the band
that forms the Fermi pocket with index $i$. The integral is performed along
each Fermi pocket; since we work with a two-dimensional model, the Fermi
pockets are closed loops $C_j$.
An eigenvalue $\lambda_\alpha\geq 1$ implies that the system is unstable
towards a SC phase with gap symmetry given by the
corresponding $\gamma_\alpha({\bf k})$. We work in the regime $T>T_c$, where all
eigenvalues are smaller than unity. Nevertheless, the symmetry of the dominant
pairing instability is given by the eigenvector to the largest eigenvalue
$\lambda_{\rm max}$.

Our calculation of the effective pairing interaction extends the one
for the single-band Hubbard model in Ref.\ \onlinecite{PhysRevB.39.11663}.
Since we only consider pairing on the Fermi surface and a
static SC gap, the interaction is assumed to be frequency independent.
We evaluate an infinite RPA-type series of
bubble and ladder diagrams. Inter-band pairing, i.e., the formation
of Cooper pairs
consisting of two electrons from different bands, either involves electrons in
states far from the Fermi energy or leads to finite-momentum Cooper pairs, and
is therefore excluded. Hence, the Cooper pairs always consist of two
electrons from the same band. The summation yields two terms that enter the
quartic part of the effective pairing Hamiltonian in addition to the bare
interaction:
\begin{widetext}
\begin{eqnarray}
H_\mathrm{pair}^\mathrm{eff} &=& -\frac{1}{N}\sum_{\nu,\mu}
  \sum_{\sigma,\sigma'} {\sum_{\bf
  k,k'}}^\prime\sum_{i\omega_n,i\omega_n'}\sum_{A,B,C,D}\sum_{j,k,n,m}\big[
  \hat{U}^z\hat{\chi}^z({\bf
  k-k'},i\omega_n-i\omega_n')\, \hat{U}^z
  \big]_{(A,B,n,\sigma),(C,D,m,\sigma')}\nonumber\\
&& {}\times \trafo{*}{A}{|j-n|}{\nu}{k'}{\sigma}\,
  \trafo{*}{C}{|k-m|}{\nu}{-k'}{\sigma'}\,
  \trafo{}{D}{k}{\mu}{-k}{\sigma'}\,
  \trafo{}{B}{j}{\mu}{k}{\sigma}\, d^\dagger\mbs{\nu}{k'}{\sigma}
  d^\dagger\mbs{\nu}{-k'}{\sigma'} d\mbs{\mu}{-k}{\sigma'}
  d\mbs{\mu}{k}{\sigma}\nonumber\\
&& {}-\frac{1}{N}\sum_{\nu,\mu}\sum_{\sigma,\sigma'} {\sum_{\bf
  k,k'}}^\prime\sum_{i\omega_n,i\omega_n'}\sum_{A,B,C,D}\sum_{j,k,n,m}\big[
  \hat{U}^{+-}\hat{\chi}^{+-}({\bf
  k-k'},i\omega_n-i\omega_n')\,\hat{U}^{+-}
  \big]_{(A,B,n,\sigma),(C,D,m,\sigma')}\delta_{\sigma,-\sigma'}\nonumber\\
&& {}\times \trafo{*}{A}{|j-n|}{\nu}{k'}{\sigma}\,
  \trafo{*}{C}{|k-m|}{\nu}{-k'}{\sigma'}\,
  \trafo{}{D}{k}{\mu}{-k}{\sigma}\,
  \trafo{}{B}{j}{\mu}{k}{\sigma'}\,
  d^\dagger\mbs{\nu}{k'}{\sigma}
  d^\dagger\mbs{\nu}{-k'}{\sigma'} d\mbs{\mu}{-k}{\sigma}
  d\mbs{\mu}{k}{\sigma'}
  \nonumber\\
&& {}+ \frac{1}{N}\sum_{\nu,\mu} \sum_{\sigma,\sigma'} {\sum_{\bf
  k,k'}}^\prime\sum_{A,B,C,D}\sum_{j,k,l} U_{(A,B),(C,D)}(\sigma,\sigma')
  \nonumber\\
&& {}\times \trafo{*}{A}{|j-l|}{\nu}{k'}{\sigma}\,
  \trafo{*}{C}{|k-l|}{\nu}{-k'}{\sigma'}\,
\trafo{}{D}{-k}{\alpha}{-k}{\sigma'}\,
  \trafo{}{B}{j}{\mu}{k}{\sigma}\, d^\dagger\mbs{\nu}{k'}{\sigma}
  d^\dagger\mbs{\nu}{-k'}{\sigma'} d\mbs{\mu}{-k}{\sigma'}
  d\mbs{\mu}{k}{\sigma}\\
&\equiv& \frac{1}{N} \sum_{\nu,\mu}\sum_{\sigma,\sigma'}{\sum_{\bf
  k,k'}}^\prime\sum_{i\omega_n,i\omega_n'}V^{z}_{\nu,\mu;\sigma,\sigma'}({\bf
  k,k'},i\omega_n-i\omega_n')\, d^\dagger\mbs{\nu}{k'}{\sigma}
  d^\dagger\mbs{\nu}{-k'}{\sigma'} d\mbs{\mu}{-k}{\sigma'} d\mbs{\mu}{k}{\sigma}
  \nonumber\\
&& {}+\frac{1}{N} 
  \sum_{\nu,\mu}\sum_{\sigma}{\sum_{\bf
  k,k'}}^\prime\sum_{i\omega_n,i\omega_n'}V^{+-}_{\nu,\mu;\sigma,-\sigma}({\bf
  k,k'},i\omega_n-i\omega_n')\, d^\dagger\mbs{\nu}{k'}{\sigma}
  d^\dagger\mbs{\nu}{-k'}{-\sigma} d\mbs{\mu}{-k}{\sigma}
  d\mbs{\mu}{k}{-\sigma} \nonumber\\
&& {}+\frac{1}{N} \sum_{\nu,\mu}\sum_{\sigma,\sigma'}{\sum_{\bf k,k'}}^\prime
  V^{0}_{\nu,\mu;\sigma,\sigma'}({\bf k,k'})\, d^\dagger\mbs{\nu}{k'}{\sigma}
  d^\dagger\mbs{\nu}{-k'}{\sigma'} d\mbs{\mu}{-k}{\sigma'}
  d\mbs{\mu}{k}{\sigma}.
\label{effpair}
\end{eqnarray}
\end{widetext}
Herein, the RPA susceptibilities take the well-known form
\begin{eqnarray}
\hat{\chi}^z({\bf q},iq_n) &=& \hat{\chi}^{z(0)}({\bf q},iq_n) \big[ \hat{1} +
  \hat{U}^z \hat{\chi}^{z(0)}({\bf q},iq_n) \big]^{-1}, \\
\hat{\chi}^{+-}({\bf q},iq_n) &=& \hat{\chi}^{+-(0)}({\bf q},iq_n) \big[
  \hat{1} - \hat{U}^{+-} \hat{\chi}^{+-(0)}({\bf q},iq_n)
  \big]^{-1} . \nonumber \\
\label{chi+-RPA}
\end{eqnarray}
The interaction matrices appearing in the effective interaction have the
components
\begin{eqnarray}
U^z_{(A,B,n,\sigma),(C,D,m,\sigma')} &=&
  U_{(A,B),(C,D)}(\sigma,\sigma')\,\delta_{n,m} \nonumber\\
&& {}- U_{(A,D),(C,B)}(\sigma,\sigma')\,\delta_{n,m}\delta_{\sigma,\sigma'},
  \nonumber \\
\\
U^{+-}_{(A,B,n,\sigma),(C,D,m,\sigma')} &=&
  U_{(A,D),(C,B)}(\sigma,\sigma')\,\delta_{n,m}\delta_{\sigma,-\sigma'}.
  \nonumber \\
\end{eqnarray}
The diagrammatic representation of the two vertices described by these
interaction matrices is shown in Figs.\ \ref{RPA}(a) and \ref{RPA}(b).
$\hat{V}^{z}$ and $\hat{V}^{+-}$ can then be understood as two separate
series of diagrams that contain either $\hat{U}^z$ or $\hat{U}^{+-}$
but are otherwise identical except for the spin indices. The two
lowest-order diagrams in these series contributing to pairing with opposite
quasi-spins are shown in Figs.\ \ref{RPA}(c) and \ref{RPA}(d).

The components of the bare susceptibility matrices are given by
\begin{eqnarray}
\lefteqn{ \chi^{z(0)}_{(A,B,n,\sigma),(C,D,m,\sigma')}({\bf q},iq_n) }
  \nonumber\\
&& \equiv \chi^{z(0)}_{(A,B,\sigma),(C,D,\sigma')}({\bf q}+n{\bf
  Q},iq_n;{\bf q}+m{\bf Q},iq_n) \nonumber\\
&& \equiv -\frac{1}{\beta V} {\sum_{{\bf k},i\omega_n}}^{\!\!\prime}
  \sum_{i,j,\nu,\mu}
  G^{(0)}_\nu({\bf k-q},i\omega_n-iq_n)G^{(0)}_\mu({\bf k},i\omega_n)
  \nonumber\\
&& {}\times\trafo{*}{A}{i}{\mu}{k}{\sigma}\,
  \trafo{}{B}{|i-n|}{\nu}{k-q}{\sigma'}\nonumber\\
&& {}\times\trafo{*}{C}{|j-m|}{\nu}{k-q}{\sigma'}\,
  \trafo{}{D}{j}{\nu}{k}{\sigma}\,
  \delta_{\sigma,\sigma'}
\label{chibare++}
\end{eqnarray}
and
\begin{eqnarray}
\lefteqn{\chi^{+-(0)}_{(A,B,n,\sigma),(C,D,m,\sigma')}({\bf q},iq_n) }
  \nonumber\\
&& \equiv \chi^{+-(0)}_{(A,B,\sigma),(C,D,\sigma')}({\bf q}+n{\bf
  Q},iq_n;{\bf q}+m{\bf Q},iq_n) \nonumber\\
&& \equiv -\frac{1}{\beta V} {\sum_{{\bf k},i\omega_n}}^{\!\!\prime}
  \sum_{i,j,\nu,\mu}
  G^{(0)}_\nu({\bf k-q},i\omega_n-iq_n)G^{(0)}_\mu({\bf k},i\omega_n)
  \nonumber\\
&& {}\times\trafo{*}{A}{i}{\mu}{k}{\sigma}\,
  \trafo{}{B}{|i-n|}{\nu}{k-q}{\sigma'}\nonumber\\
&& {}\times\trafo{*}{C}{|j-m|}{\nu}{k-q}{\sigma'}\,
  \trafo{}{D}{j}{\nu}{k}{\sigma}\,
  \delta_{\sigma,-\sigma'},
\label{chibare+-}
\end{eqnarray}
where $G^{(0)}_\nu({\bf k},i\omega_n)=(-i\omega_n + E_{\nu,{\bf k}})^{-1}$
is the bare electronic Green function in the new basis. 
The susceptibility $\hat{\chi}^{z(0)}$ describes fluctuations with spin
projection $s_z=0$ and consists of a longitudinal spin and a charge
contribution. $\hat{\chi}^{+-(0)}$ describes transverse spin fluctuations with
$s_z=\pm 1$. Note that the SDW formation does not mix states with different
$s_z$ since the \textit{z} component of spin remains conserved. Thus in this
context we do not need to distinguish between spins and quasi-spins.

The superconducting order parameter in the SDW phase is a
particle-particle expectation value of the new \textit{d} quasiparticles,
which are connected to the original electrons by the spin-dependent
transformation in Eq.\ (\ref{UTrafo}). As noted above, the quasi-spin of the
\textit{d} quasiparticles is not the same as the spin of the original electrons.
Indeed, spin-singlet and spin-triplet states with $s_z=0$ are mixed in the SDW
phase. This is clearly seen if for example the singlet order
parameter in the SDW phase is expressed in terms of the original basis:
\begin{eqnarray}
\lefteqn{ \langle d\mbs{\nu}{k}{\sigma}^\dagger d\mbs{\nu}{-k}{-\sigma}^\dagger
  - d\mbs{\nu}{k}{-\sigma}^\dagger d\mbs{\nu}{-k}{\sigma}^\dagger \rangle }
  \nonumber\\
&&= \sum_{A,m}\sum_{B,n}\big\{ {\cal U}^*_{A,m;\nu}({\bf
  k},\sigma)\,
  {\cal U}^*_{B,n;\nu}({\bf -k},-\sigma) \nonumber\\
&& {}\times \langle c^\dagger_{A,{\bf k}+m{\bf Q},\sigma}c^\dagger_{B,{\bf
  -k}+n{\bf Q},-\sigma} \rangle \nonumber\\
&& {}- {\cal U}^*_{A,m;\nu}({\bf k},-\sigma)\,
  {\cal U}^*_{B,n;\nu}({\bf -k},\sigma) \nonumber\\
&& {}\times \langle c^\dagger_{A,{\bf k}+m{\bf Q},-\sigma}c^\dagger_{B,{\bf
  -k}+n{\bf Q},\sigma} \rangle\big\}.
\end{eqnarray}
We see that an expectation value $\langle
d\mbs{\nu}{k}{\sigma}^\dagger d\mbs{\nu}{-k}{-\sigma}^\dagger -
d\mbs{\nu}{k}{-\sigma}^\dagger d\mbs{\nu}{-k}{\sigma}^\dagger\rangle$, which is
odd under quasi-spin inversion $\sigma \rightarrow-\sigma$, contains
expectation values $\langle c_{A,{\bf k}+m{\bf Q},\sigma}^\dagger c_{B,{\bf
-k}+n{\bf Q},-\sigma}^\dagger + c_{A,{\bf k}+m{\bf Q},-\sigma}^\dagger c_{B,{\bf
-k}+n{\bf Q},\sigma}^\dagger \rangle$ that are even in spin if
${\cal U}^*_{A,m;\nu}({\bf k},\sigma)\,
  {\cal U}^*_{B,n;\nu}({\bf -k},-\sigma) \neq {\cal
  U}^*_{A,m;\nu}({\bf k},-\sigma)\,
  {\cal U}^*_{B,n;\nu}({\bf -k},\sigma)$.
Analogously, a triplet order parameter can contain expectation values with
singlet symmetry in the original basis.
However, in the new basis it is still reasonable to distinguish between pairing
states that are odd in quasi-spin $\sigma$ and therefore even in ${\bf k}$ and
those that are even in $\sigma$ and odd in ${\bf k}$. In the following, we will
refer to them as quasi-spin-singlet and quasi-spin-triplet states, respectively.

Since spin-rotation symmetry is broken in the SDW phase, quasi-spin-triplet
pairing with $s_z=\pm1$ and with $s_z=0$ is not equivalent and the two cases
must be considered separately. However, the two triplet states with $|s_z|=1$
are still
degenerate. (Also recall that $s_z=\pm1$ and $s_z=0$ states are not mixed by the
SDW formation.) 

The pairing interactions in the various SC channels can be constructed from the 
effective interactions in Eq. (12). Recall that we take the pairing interactions
to be frequency independent. Hence, we take the static limit of the
susceptibilities in the following. In the static limit,
$V^{z}_{\nu,\mu;\sigma,\sigma'}$, $V^{+-}_{\nu,\mu;\sigma,\sigma'}$, and
$V^{0}_{\nu,\mu;\sigma,\sigma'}$ are symmetric under interchange of $\sigma$ and
$\sigma'$.
Therefore, we can decompose the pairing interaction,
Eq.\ (\ref{effpair}), into a singlet and two triplet channels in
the standard manner:\cite{Bickers1989206}
\begin{widetext}
\begin{eqnarray}
H^\mathrm{eff}_\mathrm{pair} &=& \frac{1}{2N} {\sum\mom{k,k'}}^\prime
  \sum_{\nu,\mu}\sum_{\sigma}\left\{
  \left[ V^{0}_{\nu,\mu;\sigma,-\sigma}({\bf k,k'}) +
  V^{z}_{\nu,\mu;\sigma,-\sigma}({\bf k,k'}) -
  V^{+-}_{\nu,\mu;\sigma,-\sigma}({\bf k,k'})\right] +
  [{\bf k}' \rightarrow -{\bf k}'] \right\} \nonumber\\
&& {}\times \big(d^\dagger\mbs{\nu}{k'}{\sigma} d^\dagger\mbs{\nu}{-k'}{-\sigma}
  - d^\dagger\mbs{\nu}{k'}{-\sigma} d^\dagger\mbs{\nu}{-k'}{\sigma}\big)
  \big(d\mbs{\mu}{-k}{-\sigma} d\mbs{\mu}{k}{\sigma} - d\mbs{\mu}{-k}{\sigma}
  d\mbs{\mu}{k}{-\sigma}\big)\nonumber\\
&& {}+ \frac{1}{2N} {\sum\mom{k,k'}}^\prime
  \sum_{\nu,\mu}\sum_{\sigma} \left\{
  \left[V^{z}_{\nu,\mu;\sigma,-\sigma}({\bf k,k'})+
  V^{+-}_{\nu,\mu;\sigma,-\sigma}({\bf k,k'})\right] -
  [{\bf k}' \rightarrow -{\bf k}'] \right\} \nonumber\\
&& {}\times \big(d^\dagger\mbs{\nu}{k'}{\sigma} d^\dagger\mbs{\nu}{-k'}{-\sigma}
  + d^\dagger\mbs{\nu}{k'}{-\sigma} d^\dagger\mbs{\nu}{-k'}{\sigma}\big)
  \big(d\mbs{\mu}{-k}{-\sigma} d\mbs{\mu}{k}{\sigma} + d\mbs{\mu}{-k}{\sigma}
  d\mbs{\mu}{k}{-\sigma}\big) \nonumber\\
&& {}+ \frac{1}{2N} {\sum\mom{k,k'}}^\prime
  \sum_{\nu,\mu}\sum_{\sigma} \left\{
  V^{z}_{\nu,\mu;\sigma,\sigma}({\bf k,k'}) - [{\bf k}'
  \rightarrow -{\bf k}'] \right\}\, d^\dagger\mbs{\nu}{k'}{\sigma}
  d^\dagger\mbs{\nu}{-k'}{\sigma} d\mbs{\mu}{-k}{\sigma} d\mbs{\mu}{k}{\sigma}
  \nonumber\\
&\equiv& \frac{1}{N} {\sum\mom{k,k'}}^\prime
  \sum_{\nu,\mu}\sum_{\sigma}
  \Gamma^{s}_{\nu,\mu}({\bf k,k'})\,
  \big(d^\dagger\mbs{\nu}{k'}{\sigma} d^\dagger\mbs{\nu}{-k'}{-\sigma} -
  d^\dagger\mbs{\nu}{k'}{-\sigma} d^\dagger\mbs{\nu}{-k'}{\sigma}\big)
  \big(d\mbs{\mu}{-k}{-\sigma} d\mbs{\mu}{k}{\sigma} - d\mbs{\mu}{-k}{\sigma}
  d\mbs{\mu}{k}{-\sigma}\big)\nonumber\\
&& {}+\frac{1}{N} {\sum\mom{k,k'}}^\prime
  \sum_{\nu,\mu}\sum_{\sigma}
  \Gamma^{t_0}_{\nu,\mu}({\bf k,k'})\,
  \big(d^\dagger\mbs{\nu}{k'}{\sigma} d^\dagger\mbs{\nu}{-k'}{-\sigma} +
  d^\dagger\mbs{\nu}{k'}{-\sigma} d^\dagger\mbs{\nu}{-k'}{\sigma}\big)
  \big(d\mbs{\mu}{-k}{-\sigma} d\mbs{\mu}{k}{\sigma} + d\mbs{\mu}{-k}{\sigma}
  d\mbs{\mu}{k}{-\sigma}\big)\nonumber\\
&& {}+ \frac{1}{N} {\sum\mom{k,k'}}^\prime
  \sum_{\nu,\mu}\sum_{\sigma}
  \Gamma^{t_1}_{\nu,\mu}({\bf k,k'})\,
  d^\dagger\mbs{\nu}{k'}{\sigma} d^\dagger\mbs{\nu}{-k'}{\sigma}
  d\mbs{\mu}{-k}{\sigma} d\mbs{\mu}{k}{\sigma},
\label{IA_s+t}
\end{eqnarray}
\end{widetext}
where $[{\bf k}' \rightarrow -{\bf k}']$ represents the preceding
terms with $\mathbf{k}'$ replaced by $-\mathbf{k}'$. 

To conclude this section we briefly comment on the relation of our approach 
to two other methods that are used to obtain effective pairing interactions from
repulsive bare interactions. The effective interactions we obtain are closely
related to the fluctuation
exchange approximation (FLEX).\cite{Bickers1989206} In the FLEX, effective
two-particle vertices are determined by taking the derivative of a generating
functional, which consists of the bare vertices and dressed Green functions,
with respect to these Green functions. If only particle-hole processes are
considered, this yields expressions for the effective interactions that have the
same form as those in Eq.\ (\ref{IA_s+t}), but with the
susceptibilities containing the dressed Green functions. In analogy to Ref.\
\onlinecite{1367-2630-11-2-025016}, our approach can be understood as an
additional approximation on top of the FLEX, consisting of replacing dressed
Green functions by bare ones. In the paramagnetic limit, our effective
interactions recover the form of the FLEX equations for a multiband
system given in Ref.\ \onlinecite{PhysRevB.69.104504}, with
dressed Green functions replaced by bare ones. Another related method is 
referred to as perturbative renormalization group (RG). Here, the diagrams that
contribute to the effective pairing interaction are only considered up to second
order and at temperature $T=0$. The condition that one eigenvalue of the gap
equation reaches unity under the RG flow yields an energy scale that is
identified with $T_c$. This method has been used to study the pairing in various
ordered phases of the single-band Hubbard model.\cite{PhysRevB.88.064505} It is
is exact in the limit of infinitesimal interactions. However, in the pnictides
the interaction strengths are of the same order as the band width so that an
approximation including higher-order diagrams is desirable.

\section{Two-band model}\label{model}

To study the effect of an effective pairing interaction mediated by spin
and charge fluctuations in a concrete multiband system, we have to
specify the band structure and the bare interactions. In the following, we will
use a two-band model that captures some important features of many iron
pnictides: There is a nearly circular hole pocket in the center of the unfolded
BZ and two approximately elliptical electron pockets around $(\pi,0)$
and $(0,\pi)$. {We divide the Hamiltonian into noninteracting and
interacting components, $H = H_{0} + H_{{\rm int}}$.
The noninteracting bands are described by
\begin{equation}
H_0 = \sum_{\bf k}\sum_{\sigma}
  \big( \epsilon^{c}_{\bf k}c^{\dagger}_{{\bf
    k}\sigma}c^{}_{{\bf k}\sigma} + \epsilon^{f}_{\bf k}
    f^{\dagger}_{{\bf k}\sigma}f^{}_{{\bf k}\sigma} \big) ,
\end{equation}
where $c^\dagger_{{\bf k}\sigma}$ ($f^\dagger_{{\bf k}\sigma}$) creates a
spin-$\sigma$ electron with momentum ${\bf k}$ in the hole-like (electron-like)
band. Neglecting the small orthorhombic distortion, the
dispersions are\cite{PhysRevB.84.214510}
$\epsilon\mom{k}^c=\epsilon_c +
2t_c\,(\cos k_xa + \cos k_ya) - \mu$ and $\epsilon\mom{k}^f = \epsilon_f +
4t_f\cos k_xa \cos k_ya - t_f\xi_e\,(\cos k_xa + \cos k_ya) - \mu$, where $a$
is the Fe-Fe bond length and $\mu$ is the chemical potential. In units of $t_c$
we set $t_f=t_c$, $\epsilon_c=-3.5t_c$, and $\epsilon_f=3.0t_c$. The parameter
$\xi_e$ determines the ellipticity of the electron pockets. Here, we choose
$\xi_e=1$, which corresponds to moderate ellipticity.
Figure \ref{FermiS}(a) shows the resulting Fermi surface for an electron
doping level of $\delta n=0.085$ relative to half filling. 

\begin{figure}[b]
\includegraphics[clip,width=\columnwidth]{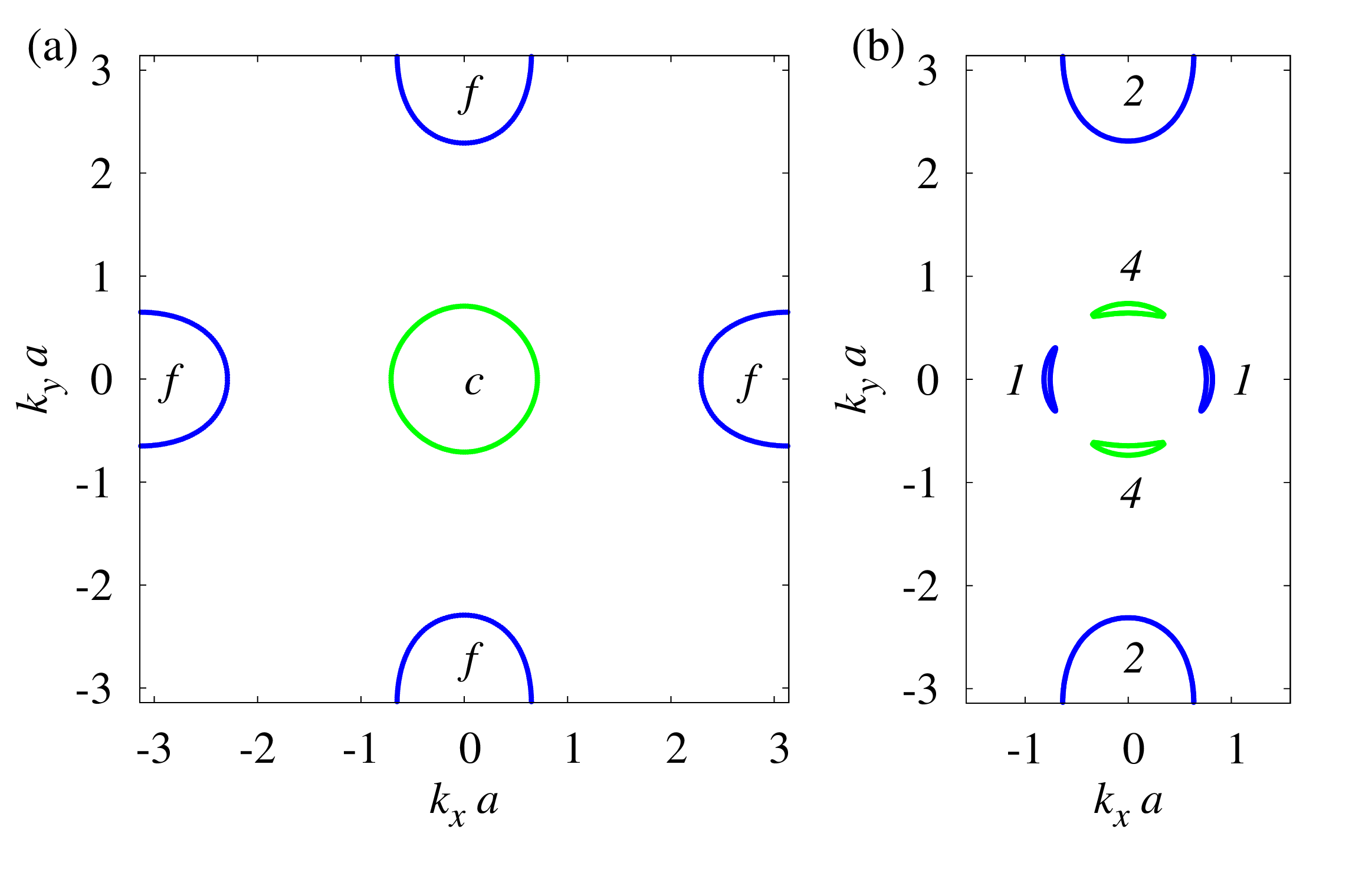}
\caption{Fermi surface for our model (a) in the paramagnetic phase and (b) in
the SDW phase with a SDW gap of $\Delta=0.055 t_c$ which corresponds to a
temperature of $k_BT=0.06 t_c$. The doping has been
set to $\delta n = 0.085$. In the paramagnetic phase, there is one hole pocket
(light green/light gray) and two electron pockets (blue/dark gray). In the SDW
phase, there are two hole pockets (light green/light gray) and three electron
pockets (blue/dark gray). The letters and numbers specify the bands that form
the corresponding Fermi pockets.}
\label{FermiS}
\end{figure}

Following Ref.\ \onlinecite{PhysRevB.78.134512}, we  include four on-site
interaction terms in $H_{{\rm int}}$: the intraband Coulomb repulsion,
which we set to be equal for both bands,
\begin{eqnarray}
H_{1} &=& \frac{g_1}{V}\sum\mom{k,k',q} \big( c^\dagger_{{\bf
  k+q}\uparrow}c^\dagger_{{\bf k'-q}\downarrow}c_{{\bf k'}\downarrow}c_{{\bf
  k}\uparrow} \nonumber\\
&& {}+ f^\dagger_{{\bf k+q}\uparrow}f^\dagger_{{\bf
  k'-q}\downarrow}f_{{\bf
  k'}\downarrow}f_{{\bf k}\uparrow} \big),
\end{eqnarray}
the interband Coulomb repulsion
\begin{equation}
H_{cf} = \frac{g_{cf}}{V}\sum\mom{k,k',q}\sum_{\sigma,\sigma'} c^\dagger_{{\bf
  k+q}\sigma}f^\dagger_{{\bf k'-q}\sigma'}f_{{\bf k'}\sigma'}c_{{\bf k}\sigma},
\end{equation}
and two types of correlated interband-hopping transitions,
\begin{eqnarray}
H_{2a} &=& \frac{g_{2a}}{V}\sum\mom{k,k',q} \big( c^\dagger_{{\bf
  k+q}\uparrow}c^\dagger_{{\bf k'-q}\downarrow}f_{{\bf k'}\downarrow}f_{{\bf
  k}\uparrow} + {\rm H.c.} \big),\\
H_{2b} &=& \frac{g_{2b}}{V}\sum\mom{k,k',q}\sum_{\sigma,\sigma'}c^\dagger_{{\bf
  k+q}\sigma}f^\dagger_{{\bf k'-q}\sigma'}c_{{\bf k'}\sigma'}f_{{\bf
  k}\sigma}.
\label{g_2b}
\end{eqnarray}
With these interactions, the interaction matrix in Eq.\ (\ref{Sint}) takes the
form
\begin{eqnarray}
\lefteqn{ \hat{U}(\sigma,\sigma') } \nonumber\\
&& = \frac{1}{V}\left(
  \begin{array}{*{4}{c}}
    g_1\delta_{\sigma,-\sigma'} & 0 & 0 & g_{cf} \\
    0 & g_{2a}\delta_{\sigma,-\sigma'} & g_{2b} & 0 \\
    0 & g_{2b} & g_{2a}\delta_{\sigma,-\sigma'} & 0 \\
    g_{cf} & 0 & 0 & g_1\delta_{\sigma,-\sigma'}\\
  \end{array} \right).\quad
\label{U2band}
\end{eqnarray}
Only two of the interactions are responsible for the formation of a
SDW gap so that a SDW interaction strength $g_{\rm SDW} \equiv g_{cf}
+ g_{2a}$ can be defined.~\cite{PhysRevB.24.5713} 

We have previously shown that this model exhibits
a robust SDW phase with
ordering vector $\mathbf{Q}=(\pi,0)$ or $(0,\pi)$ for an extended doping range
around $\delta n=0.085$.\cite{PhysRevB.84.214510} Decoupling within a
mean-field approximation, we obtain the Hamiltonian
\begin{eqnarray}
H_{\rm MF} &=& {\sum_{{\bf k},\sigma}}' \left( c^\dagger_{{\bf k}\sigma},
  c^\dagger_{{\bf k+Q}\sigma}, f^\dagger_{{\bf k}\sigma}, f^\dagger_{{\bf
  k+Q}\sigma} \right) \nonumber\\
&& {}\times\left(
  \begin{array}{*{4}{c}}
    \epsilon\mom{k}^c & 0 & 0 & \sigma\Delta \\
    0 & \epsilon\mom{k+Q}^c & \sigma\Delta & 0 \\
    0 & \sigma\Delta & \epsilon\mom{k}^f & 0 \\
    \sigma\Delta & 0 & 0 & \epsilon\mom{k+Q}^f\\
  \end{array} \right) \left(
  \begin{array}{*{1}{c}}
    c_{{\bf k}\sigma}\\
    c_{{\bf k+Q}\sigma}\\
    f_{{\bf k}\sigma}\\
    f_{{\bf k+Q}\sigma}
    \end{array}
  \right),\qquad
\label{H2free}
\end{eqnarray}
where
\begin{equation}
\Delta = -\frac{g_{\rm SDW}}{2V}\sum_{{\bf k},\sigma}\sigma\,
  \langle f^\dagger_{{\bf k}\sigma}c_{{\bf k+Q}\sigma} \rangle_{\rm MF}
\end{equation}
is the mean-field SDW gap and $\langle \ldots \rangle_{\rm MF}$ indicates the
thermal average calculated with $H_{\rm MF}$. The mean-field Hamiltonian is
diagonalized by a unitary matrix of the form
\begin{equation}
\hat{\cal U}({\bf k},\sigma) = \left(
  \begin{array}{*{4}{c}}
    -\sigma u_{1,{\bf k}} & 0 & 0 & -v_{1,{\bf k}} \\
    0 & -\sigma u_{2,{\bf k}} & -v_{2,{\bf k}} & 0 \\
    0 & v_{2,{\bf k}} & -\sigma u_{2,{\bf k}} & 0 \\
    v_{1,{\bf k}} & 0 & 0 & -\sigma u_{1,{\bf k}}\\
  \end{array} \right) ,
\label{2bTrafo}
\end{equation}
with the transformation factors
\begin{eqnarray}
u_{1,{\bf k}} &=& \frac{\epsilon_{fc}^-({\bf k}) - \sqrt{\epsilon_{fc}^-({\bf
  k})^2 + \Delta^2}}{\sqrt{\Delta^2 + \left(\epsilon_{fc}^-({\bf k}) -
  \sqrt{\epsilon_{fc}^-({\bf k})^2 + \Delta^2}\right)^2}},\\
v_{1,{\bf k}} &=& \frac{\Delta}{\sqrt{\Delta^2 +\left(\epsilon_{fc}^-({\bf k}) -
  \sqrt{\epsilon_{fc}^-({\bf k})^2 + \Delta^2}\right)^2}},
\end{eqnarray}
and $u_{2,{\bf k}} = u_{1,{\bf k+Q}}$, $v_{2,{\bf k}} = v_{1,{\bf k+Q}}$. The
reconstructed bands are given by 
\begin{eqnarray}
E_1(\mathbf{k}) &=& \epsilon_{fc}^+(\mathbf{k}) +
  \sqrt{\epsilon_{fc}^-(\mathbf{k})^2 + \Delta^2} , \\
E_2(\mathbf{k}) &=& \epsilon_{cf}^+(\mathbf{k}) +
  \sqrt{\epsilon_{cf}^-(\mathbf{k})^2 + \Delta^2} , \\
E_3(\mathbf{k}) &=& \epsilon_{cf}^+(\mathbf{k}) -
  \sqrt{\epsilon_{cf}^-(\mathbf{k})^2 + \Delta^2} , \\
E_4(\mathbf{k}) &=& \epsilon_{fc}^+(\mathbf{k}) -
  \sqrt{\epsilon_{fc}^-(\mathbf{k})^2 + \Delta^2} ,
\end{eqnarray}
where $\epsilon_{ij}^\pm({\bf k}) = (\epsilon^i_{{\bf k+Q}}\pm
  \epsilon^j_{{\bf k}})/2$ and
$i$ and $j$ can be $c$ or $f$.
The resulting reconstructed Fermi surface is shown in Fig.\ \ref{FermiS}(b). In
the SDW phase, the hole Fermi pocket and the electron pocket around
$\mathbf{Q}=(\pi,0)$, which is strongly nested with the hole pocket,
reconstruct to form four small banana-shaped pockets. Two of these are
electron-like and two are hole-like. The electron pocket around $(0,\pi)$ is
only weakly affected by the SDW.}

\section{Analysis of the effective interaction}\label{analytical}

Even the minimal model introduced above cannot be solved analytically. The
matrices $\hat{\chi}^{z}$, $\hat{\chi}^{+-}$, $\hat{\Gamma}^{s}$,
$\hat{\Gamma}^{t_0}$, and $\hat{\Gamma}^{t_1}$ each contain $16\times 16$
components
and for non-parabolic bands it is impossible to analytically
calculate the susceptibilities appearing in the effective interactions.
Nevertheless, it is possible to draw some conclusions about the effective
interactions based on analytical considerations, which helps to understand the
numerical results presented in Sec.\ \ref{numerics}.

{The presence of the Goldstone magnon mode in the SDW phase implies
divergent static transverse spin suscpetibilities. Specifically, the
components $\chi_{(c,f,1,\uparrow),(c,f,1,\downarrow)}^{+-}$,
$\chi_{(c,f,1,\uparrow),(f,c,1,\downarrow)}^{+-}$,
$\chi_{(f,c,1,\uparrow),(c,f,1,\downarrow)}^{+-}$, and
$\chi_{(f,c,1,\uparrow),(f,c,1,\downarrow)}^{+-}$, and their sum,
diverge for ${\bf q}\to 0$  in the magnetic BZ. 
This begs the question of whether these components lead to a singular
contribution to the effective pairing interaction. We first note that
this {question only pertains to
the singlet and $s_{z}=0$ triplet pairing
interactions, $\Gamma^{s}({\bf k}, {\bf k}')$ and $\Gamma^{t_0}({\bf
  k},{\bf k}')$ respectively, as only
these terms include the contribution
$V^{+-}_{\nu,\mu;\sigma,-\sigma}({\bf k-k'})$ 
from the transverse susceptibilities. Furthermore, a possible
divergence of these interactions can only occur at ${\bf k}=\pm{\bf
  k}'$. At these points the contribution of the divergent
susceptibilities to the interaction is proportional to
\begin{eqnarray}
\sum_{A\neq B}\lefteqn{\Big\{ \big[\hat{U}^{+-}\hat{\chi}^{+-}({0})\,
  \hat{U}^{+-}\big]_{(A,B,1,\downarrow),(A,B,1,\uparrow)}} \nonumber\\
&& - \big[\hat{U}^{+-}\hat{\chi}^{+-}({0})\,
  \hat{U}^{+-}\big]_{(A,B,1,\downarrow),(B,A,1,\uparrow)}
\Big\}.\label{diverging_diff}
\end{eqnarray}
For the special case $g_{cf}\neq0$ and $g_1=g_{2a}=g_{2b}=0$, this is
in turn proportional to the difference
$\chi^{+-}_{(A,B,1,\uparrow),(A,B,1,\downarrow)}(0) 
  -\chi^{+-}_{(A,B,1,\uparrow),(B,A,1,\downarrow)}(0)$, where $A\neq B$.
Using the RPA equations (\ref{chi+-RPA}), this difference can be
rewritten as
\begin{eqnarray}
\lefteqn{ \chi^{+-}_{(A,B,1,\uparrow),(A,B,1,\downarrow)}(0)-\chi^{+-}_{(A,B,1,
  \uparrow),(B,A,1,\downarrow)}(0) } \nonumber\\
&& = \frac{\chi^{+-(0)}_{(A,B,1,\uparrow),(A,B,1,\downarrow)}(0)
  -\chi^{+-(0)}_{(A,B,1,\uparrow),(B,A,1,\downarrow)}(0)}{1-g_{cf}\big[
  \chi^{+-(0)}_{(A,B,1,\uparrow),(A,B,1,\downarrow)}(0)-\chi^{+-(0)}_{(A,B,1,
  \uparrow),(B,A,1,\downarrow)}(0) \big]} . \nonumber \\
\label{chi_diff}
\end{eqnarray}
The denominator of Eq.\ (\ref{chi_diff}) is non-zero, however, as the
individual interband susceptibilities and their sum diverge if
\begin{equation}
\chi^{+-(0)}_{(A,B,1,\uparrow),(B,A,1,\downarrow)}(0) +
  \chi^{+-(0)}_{(A,B,1,\uparrow),(A,B,1,\downarrow)}(0)=\frac{1}{g_{cf}}.
\label{div-cond}
\end{equation}
Since the denominator contains the difference instead of the sum, we hence 
conclude that the contribution of Eq.~(\ref{diverging_diff}) to
the effective interaction remains finite. We note that a
non-vanishing contribution to the interaction does not violate
Adler's theorem, which states that the vertex function describing the 
coupling of electrons to a Goldstone mode vanishes for zero tranferred
momentum,\cite{adler, adlers} since a divergence of the magnon propagator
compensates for the vanishing vertex function. A similar compensation has been
found for the single-band Hubbard model applied to
cuprates.\cite{PhysRevB.39.11663, PhysRevB.46.11884, chubukov_pines} 

In the general case where all of the interaction potentials are
allowed to be
non-zero, we have found numerically that the pairing interaction
remains finite at ${\bf k=k'}$ and is a smooth function of the
momenta. In Fig.\ \ref{V+-_alpha}, we plot
$V^{+-}_{\nu,\mu;\sigma,-\sigma}({\bf 
  k-k'})$ for ${\bf k}$ close to ${\bf k}'$ and ${\bf k}'$ lying
on one of the banana-shaped electron pockets for
various combinations of the interaction parameters. We see
that the effective interaction is indeed a smooth function of momentum. This
also justifies dropping the exclusion of the point
${\bf q}={\bf Q}$ from the momentum sums in Sec.\ \ref{saddle-pt}. }

\begin{figure}
\includegraphics[clip, width=\columnwidth]{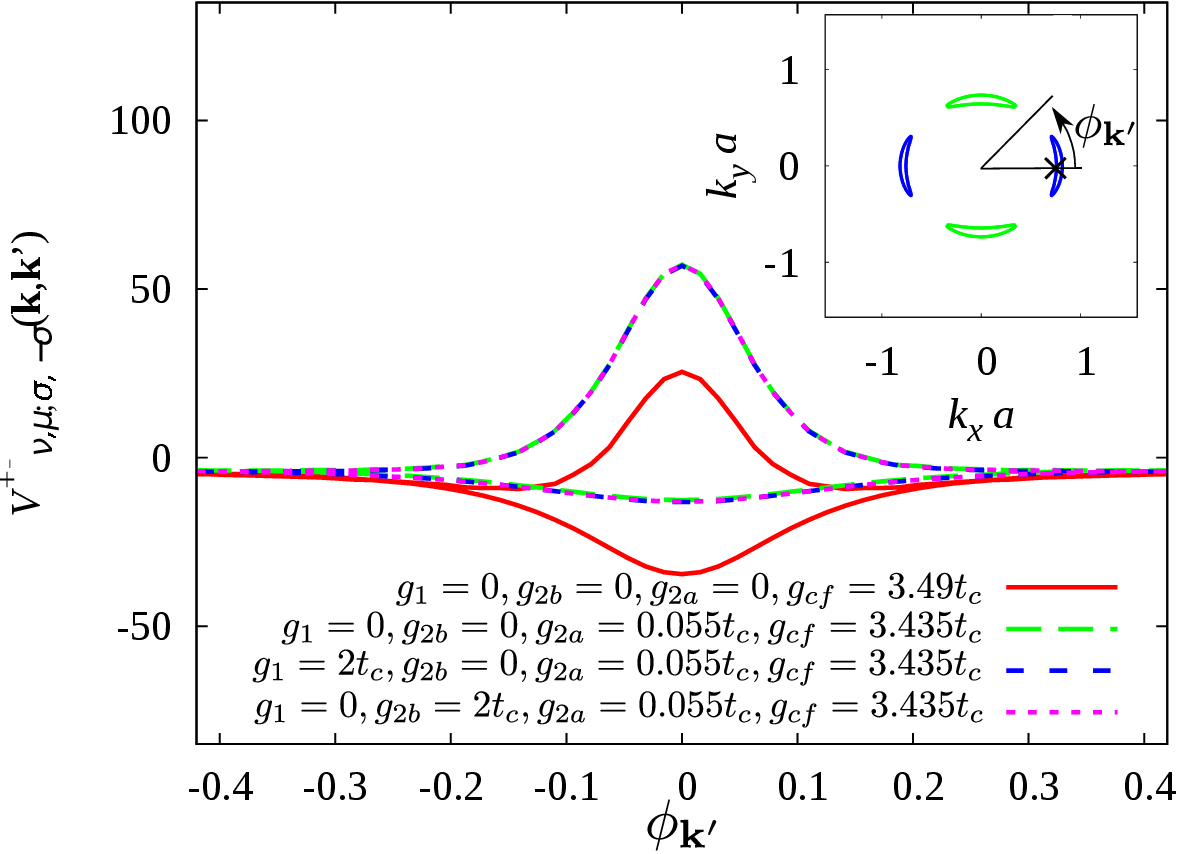}
\caption{Effective pairing interaction
$V^{+-}_{\nu,\mu;\sigma,-\sigma}({\bf k-k'})$ around the banana-shaped
  electron pocket due to
transverse spin fluctuations for small ${\bf k}-{\bf k}'$ as a function of the
polar angle $\phi_{\mathbf{k}'}$ of ${\bf k}'$ which is
  shown in the inset. The second momentum ${\bf 
k}$ is indicated by a black cross.  We show two curves for each
combination of the bare interaction strengths, one corresponding to the
inner and the other to the outer part of the pocket.}
\label{V+-_alpha}
\end{figure}

\section{Numerical Results}\label{numerics}

In this section, we present numerical results for the SC gap structure and its
dependence on the interactions $g_{cf}$, $g_1$, $g_{2a}$, and $g_{2b}$. For the
numerical solution of the mean-field equations for the SDW order parameters
$\Delta$, we use a $400\times400$ {\bf k}-point mesh in the paramagnetic
BZ.  The calculation of the bare susceptibilities is performed
using a
$100\times100$ {\bf k}-point mesh. Finally, to solve the SC gap
equation~(\ref{gapeq}) we discretize the Fermi surface into 
158 points. 128 points of these are chosen on the small banana-shaped
Fermi pockets because the calculations are much more sensitive to changes in
the number of {\bf k}-points on these strongly reconstructed pockets.
The doping is chosen as $\delta n = 0.085$. The SDW interaction
is set to $g_{\rm SDW}=3.49t_c$, which gives an ordering temperature of $k_B
T_N\approx 0.065t_c$ and a reasonable ratio of the zero temperature SDW gap to
the band width.\cite{PhysRevB.84.214510} The effective pairing
interaction is calculated for a temperature of $k_B T=0.06t_c$.

\subsection{Interband Coulomb repulsion $g_{cf}$ and interband hopping
$g_{2a}$}\label{sec_IA_g1_0}

\begin{figure}
\includegraphics[clip,width=\columnwidth]{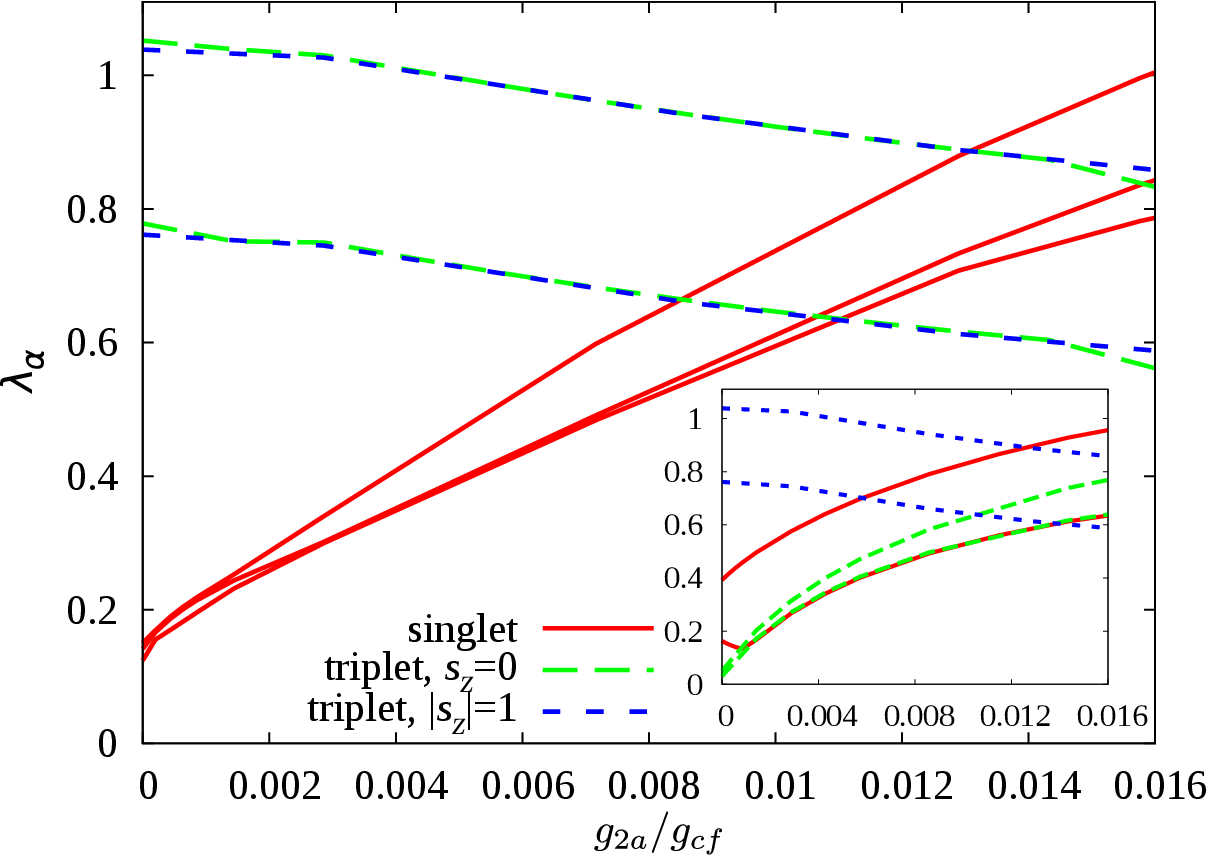}
\caption{Largest eigenvalues of the linearized gap equation in
the singlet and triplet channels, as functions of $g_{2a}/g_{cf}$. We have set
$g_1 = g_{2b} = 0$. The inset shows the largest eigenvalues obtained if only 
longitudinal fluctuations and the bare interaction are considered.}
\label{lambda_vs_g3}
\end{figure}

\begin{figure}
\includegraphics[clip,width=\columnwidth]{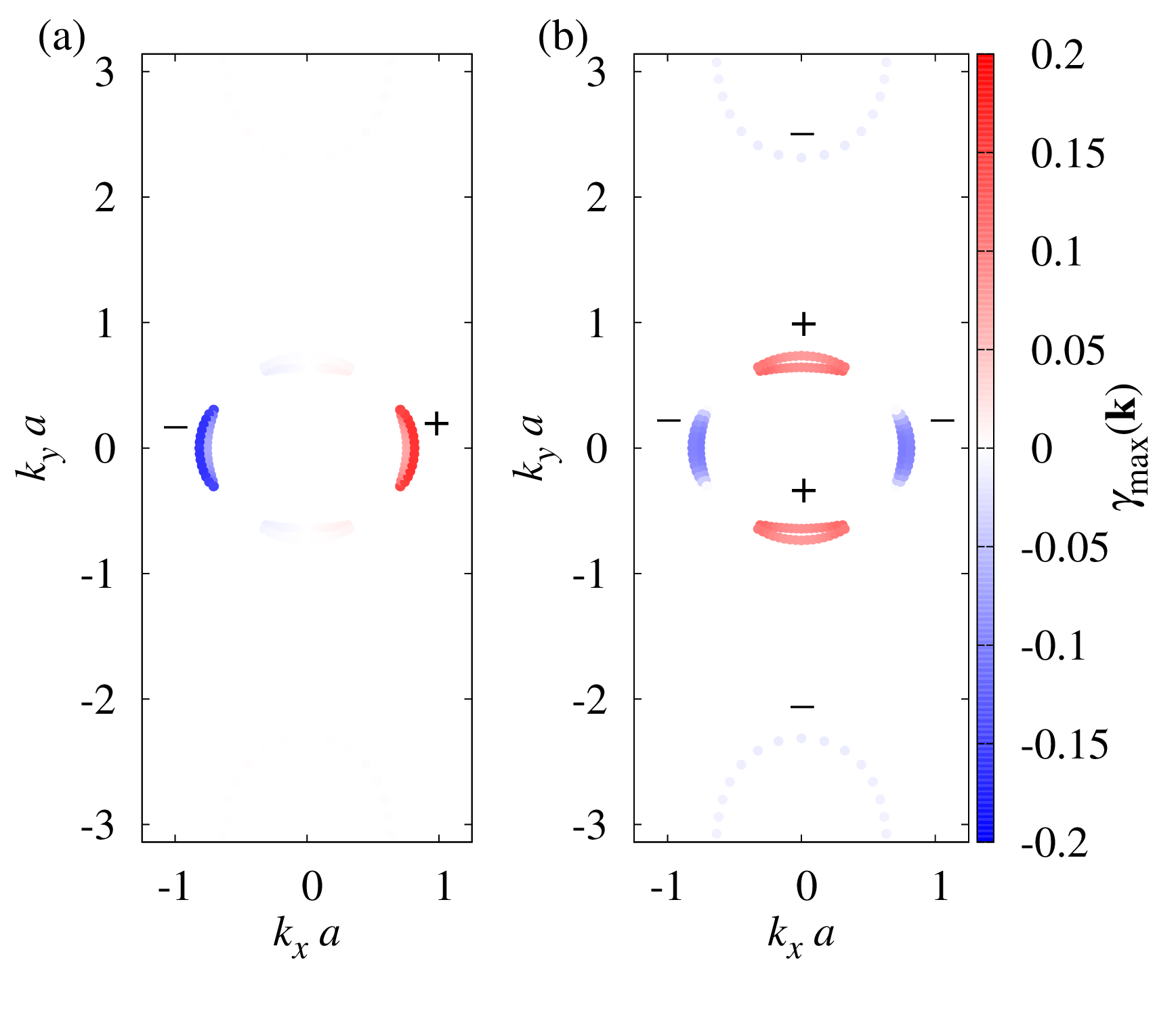}
\caption{Gap structure of (a) the dominant $s_z=0$ triplet state and (b) the
dominant singlet state for $g_{2a}/g_{cf}=0.016$. We have set $g_1 = g_{2b} =
0$.}
\label{gap_g3}
\end{figure}

\begin{figure*}
\includegraphics[clip,width=0.75\textwidth]{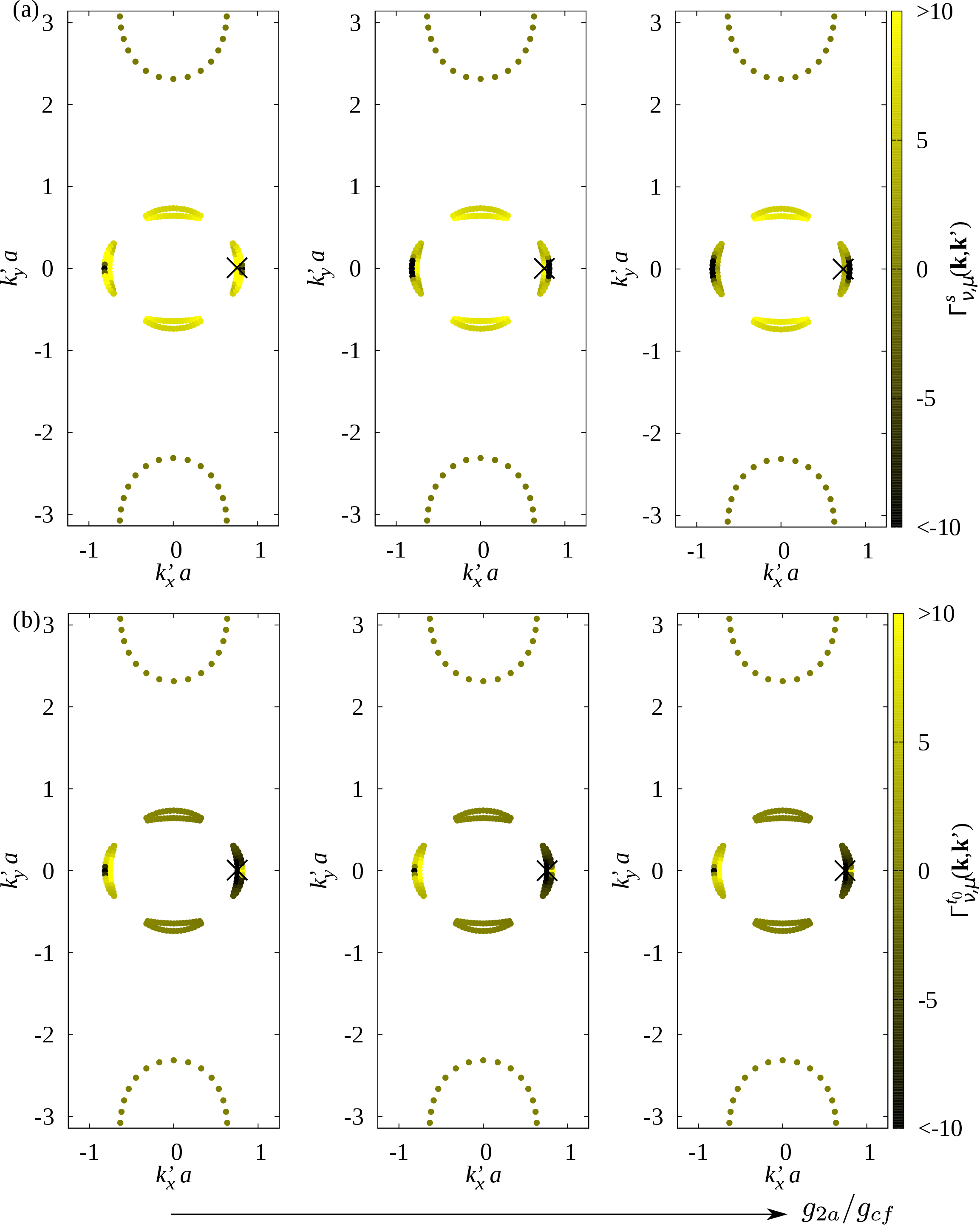}
\caption{Effective pairing interaction on the Fermi surface as a
function of ${\bf k'}$. The value of the momentum ${\bf k}$ is indicated by the
black cross and the first band index is set to $\nu=1$. From left to right we
plot the interaction on the Fermi surface for $g_{2a}/g_{cf}=0$,
$g_{2a}/g_{cf}=0.009$, and $g_{2a}/g_{cf}=0.016$ in (a) the singlet channel and
(b) the $s_z=0$ triplet channel. We have set $g_1 =g_{2b} = 0$.}
\label{IA_g2a}
\end{figure*}

We first discuss the case where the interactions that do not support a SDW
vanish,  and so we set $g_1 = g_{2b} = 0$ while fixing the sum of
  the Coulomb repulsion $g_{cf}$ and the pair-hopping amplitude
  $g_{2a}$ to be $g_{cf}+g_{2a}=g_{\rm
  SDW}=3.49t_c$. Since a negative value of
$g_{2a}$ leads to a charge-density-wave instead of a SDW
state,\cite{PhysRevB.80.174401,PhysRevB.24.5713} we only consider $g_{2a}\ge0$.

In Fig.\ \ref{lambda_vs_g3}, we plot the largest eigenvalues obtained from the
SC gap equation (\ref{gapeq}) in the quasi-spin singlet and triplet channels, as
functions of the ratio $g_{2a}/g_{cf}$.  For very small pair-hopping
amplitudes, the triplet
pairing dominates. Although the strict degeneracy of the triplet states with
$s_z=0$ and $s_z=1$ is broken, they are nearly degenerate over the complete
parameter range and show the same gap structure. The gap structure of the
leading
triplet state is shown in Fig.\ \ref{gap_g3}(a). It has the symmetry of a
$p_x$-wave state with most of the gap weight on the small electron pockets.
Upon increasing the ratio $g_{2a}/g_{cf}$, the eigenvalues belonging to the
$s_z=0$ triplet states decrease, while the eigenvalues for the singlet states
increase. At $g_{2a}/g_{cf}\approx 0.013$, a singlet state becomes the
leading SC instability. The gap structure of the leading singlet state
is shown in Fig.\ \ref{gap_g3}(b);
it has the structure of the $s^\pm$-type state predicted
earlier.\cite{PhysRevB.81.174538, PhysRevB.82.014521,
  PhysRevB.84.180510} Below $g_{2a}/g_{cf}\approx 0.004$ and above 
  $g_{2a}/g_{cf}\approx 0.016$, the largest
eigenvalue exceeds unity. This means that the system becomes
unstable towards a SC state. While this formally contradicts the
assumption of a normal conducting state made in the derivation, the eigenvector
to the largest eigenvalue still gives a good indication of the leading
instability.  Since the predicted SC critical temperature
becomes much higher than experimentally observed values
for $g_{2a}/g_{cf}\gtrsim 0.016$, we exclude this parameter range.

The inset in Fig.\ \ref{lambda_vs_g3} shows the evolution of the largest 
eigenvalues as functions of $g_{2a}/g_{cf}$ when the transverse contribution to
the interaction is set to zero. In this case the $s_z=0$ triplet pairing channel
is most strongly reduced while the $s_z=1$ triplet is completely unaffected
because it originates only from the longitudinal fluctuations. The singlet
channel lies in between these extremes. If only the bare interactions are
considered the eigenvalues in the triplet channels are strictly zero while the
largest eigenvalue in the singlet channel is proportional to $g_{2a}/g_{cf}$ and
is reduced by a factor of about $10^{-2}$ compared to the calculation with the
full interaction. This shows that the spin and charge fluctuations strongly
promote the pairing in the SDW phase.

The crossover from $p_x$-wave to $s^\pm$-wave pairing can be
understood from the evolution of the effective pairing interaction with
$g_{2a}/g_{cf}$, which is shown in Fig.\ \ref{IA_g2a} as a function of
$\mathbf{k}'$, for $\mathbf{k}$ lying on the inner part of the right
banana-shaped electron pocket. The
interaction is peaked at ${\bf k}'=\pm {\bf k}$. This peak extends
to the other side of the banana-shaped electron pocket, where it takes
 the opposite sign due to the SDW
transformation factors multiplying the susceptibilities. The peak appears in
the transverse contribution to the pairing interaction and 
therefore enters with  opposite signs in the singlet and $s_z=0$
triplet channels, see Eq.\ (\ref{IA_s+t}). For
$g_{2a}/g_{cf}=0$, the peak is strongly negative (positive) and
therefore attractive (repulsive) for ${\bf k}'\approx {\bf k}$
(${\bf k}'\approx - {\bf k}$) in the triplet channel, which supports a sign
change of the gap under ${\bf k} \rightarrow -{\bf k}$ and therefore favors a
\textit{p}-wave state. At the same time, the interaction with the other Fermi
pockets is weak and thus does not suppress the \textit{p}-wave
state. Upon increasing $g_{2a}/g_{cf}$, the repulsive peak in the
singlet  interaction at $\mathbf{k}'\approx\pm\mathbf{k}$ is suppressed,
 while the attractive interaction for
$\mathbf{k}'$ on the other (outer) side of the banana-shaped electron pocket
remains strong, see Fig.\ \ref{IA_g2a}(a). Overall, this leads to a
stronger attractive pairing interaction between the two small electron
pockets, which favors a singlet state. The repulsion between the small
electron and hole pockets then  stabilizes a $s^\pm$-type
structure. In contrast, there is little change in the form of the
triplet interaction with increasing $g_{2a}/g_{cf}$, although the
strength is overall slightly reduced, see
Fig.~\ref{IA_g2a}(b).

In Fig.\ \ref{lambda_vs_g3}, we also plot smaller eigenvalues in each channel.
In the $s_z=0$ triplet channel, the second largest eigenvalue is clearly
separated from the largest eigenvalue and corresponds to a $p_y$-wave gap with a
line node along the $k_x$ axis. In the singlet channel, the
three largest eigenvalues are nearly degenerate for $g_{2a}/g_{cf}=0$, but at
finite $g_{2a}/g_{cf}\approx 0.004$ they split up.
The second and third eigenvalues are nearly degenerate for the interval
$0.004\lesssim g_{2a}/g_{cf}\lesssim 0.011$. For larger $g_{2a}/g_{cf}$,
the second largest eigenvalue has a $d_{xy}$-type structure with nodes along the
$k_x$ and $k_y$ axes.

\subsection{Intraband Coulomb repulsion $g_1$}\label{intraband_coul}

\begin{figure}
\includegraphics[clip,width=\columnwidth]{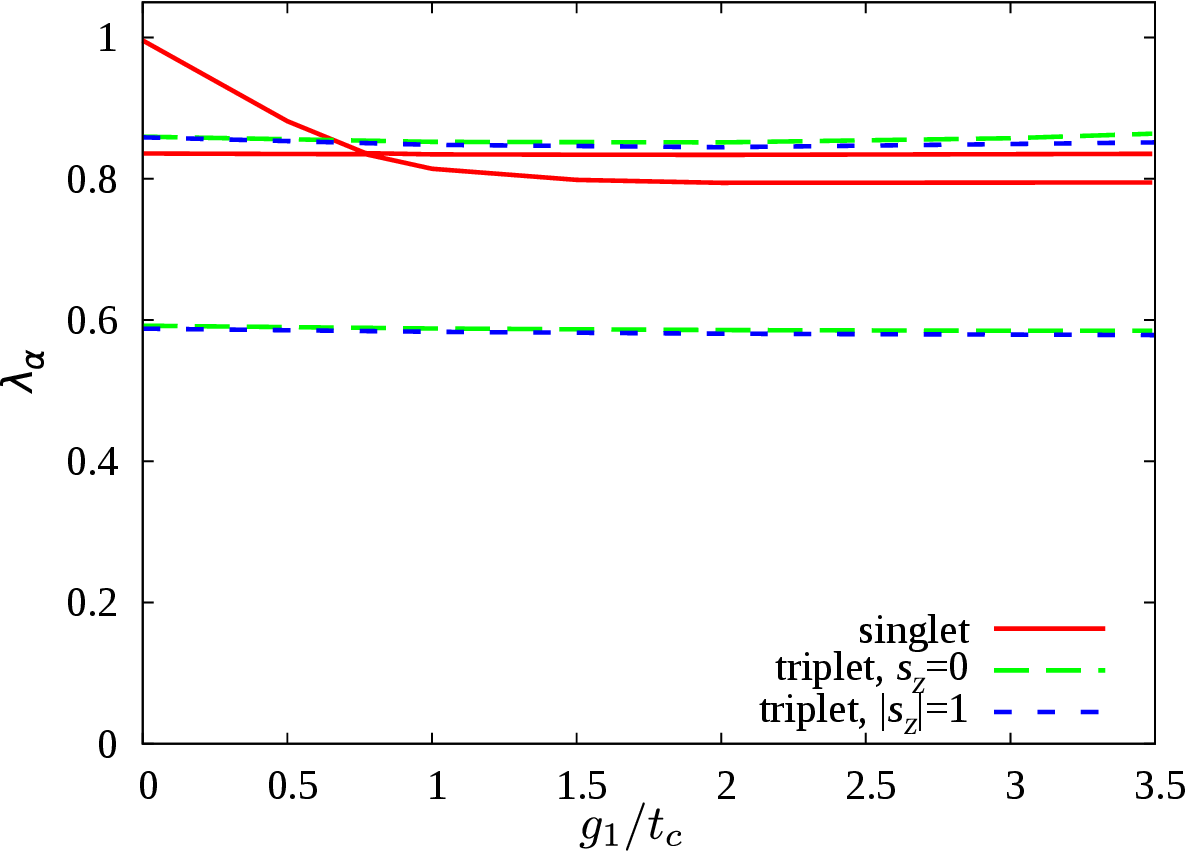}
\caption{Largest eigenvalues of the linearized gap equation, as functions 
of the intraband Coulomb interaction $g_1$.
The ratio of the pair hopping and
the interband repulsion has been set to $g_{2a}/g_{cf}=0.016$ with $g_{\rm
SDW}=3.49 t_c$ and $g_{2b}=0$.}
\label{lambda_vs_g1}
\end{figure}

We next discuss the intraband Coulomb repulsion with interaction strength
$g_1$. This term does not affect the SDW order at the mean-field level
but can change the SC pairing. We choose the ratio $g_{2a}/g_{cf}=0.016$, for
which we have found a $s^{\pm}$-type singlet state as the leading SC
instability, and set $g_{2b}=0$. According to Ref.\
\onlinecite{PhysRevB.78.134512}, $g_1\approx g_{cf}$ holds if the
electron and hole pockets have the same shape. Since we
assume weakly elliptical electron pockets we allow for a slightly larger $g_1$
and restrict ourselves to the range of $0\le g_1\leq 3.5t_c$ in the following.

\begin{figure}
\includegraphics[clip,width=\columnwidth]{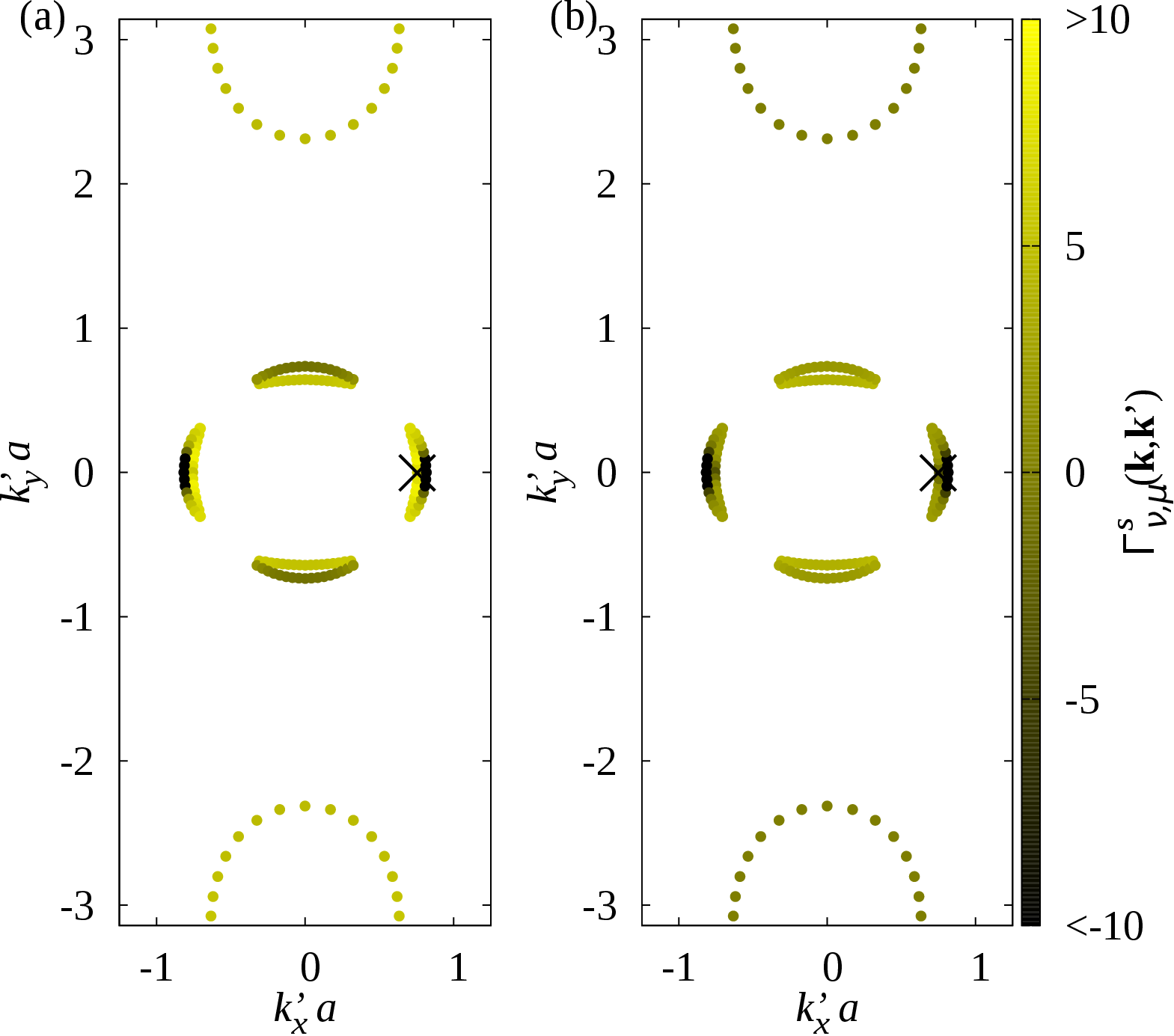}
\caption{Effective pairing interaction on the Fermi surface in the singlet
channel as a function of ${\bf k'}$ for (a) $g_{2a}/g_{cf}=0.016$,
$g_1=2 t_c$, and $g_{2b}=0$ and (b) $g_{2a}/g_{cf}=0.016$, $g_1=0$, and
$g_{2b}=2 t_c$. The
value of momentum ${\bf k}$ is indicated by the black cross and the first band
index is set to $\nu=1$.}
\label{IA_g1}
\end{figure}

In Fig.\ \ref{lambda_vs_g1} we plot the largest eigenvalues from the SC gap
equation in the singlet and triplet channels as functions of $g_1$. The
figure shows that a finite $g_1$ leads to the suppression of
singlet pairing, while the quasi-spin-triplet states are hardly affected.
Consequently, at $g_1\approx
0.1 t_c$ the triplet states become the dominant pairing instabilities
again. The gap structure of the dominant triplet state is still the
$p_x$-wave depicted in Fig.\ \ref{gap_g3}(a). The suppression of the
singlet state can be understood from the interaction in the singlet channel,
which we plot in Fig.\ \ref{IA_g1}(a) for $g_1=2 t_c$: The intrapocket
interaction is enhanced by the finite $g_1$ and the interaction between the
electron and
the hole pockets becomes less strongly repulsive. Also, the interaction between
the small electron pocket and the large electron pocket becomes weakly
repulsive. These tendencies disfavor a sign change of the SC gap between
electron and hole pockets and hence suppress the eigenvalue
corresponding to $s^\pm$-type pairing.

The intraband Coulomb repulsion also significantly modifies the dominant
singlet pairing state. Already for moderate $g_1\approx 0.5t_c$,
the $s^{\pm}$-type state develops accidental nodes on the small electron
pockets, as shown in Fig.\ \ref{gap_g1-g3}(a).
At $g_1\approx 0.7 t_c$, the two largest eigenvalues
in the quasi-spin-singlet channel cross and a state with
nodes along the $k_x$ and $k_y$ axes 
becomes the dominant singlet state. The gap is plotted in 
Fig.\ \ref{gap_g1-g3}(b). After the crossing, when the
$s^\pm$-type state is subdominant, it assumes the structure shown in
Fig.\ \ref{gap_g1-g3}(c). 
The two largest eigenvalues in the singlet channel remain very close to each
other up to $g_1=3.5 t_c$. The appearance of nodes in the gap can be
attributed to the increase of the intraband repulsion
seen in Fig.\ \ref{IA_g1}.

\begin{figure*}
\includegraphics[clip,width=0.75\textwidth]{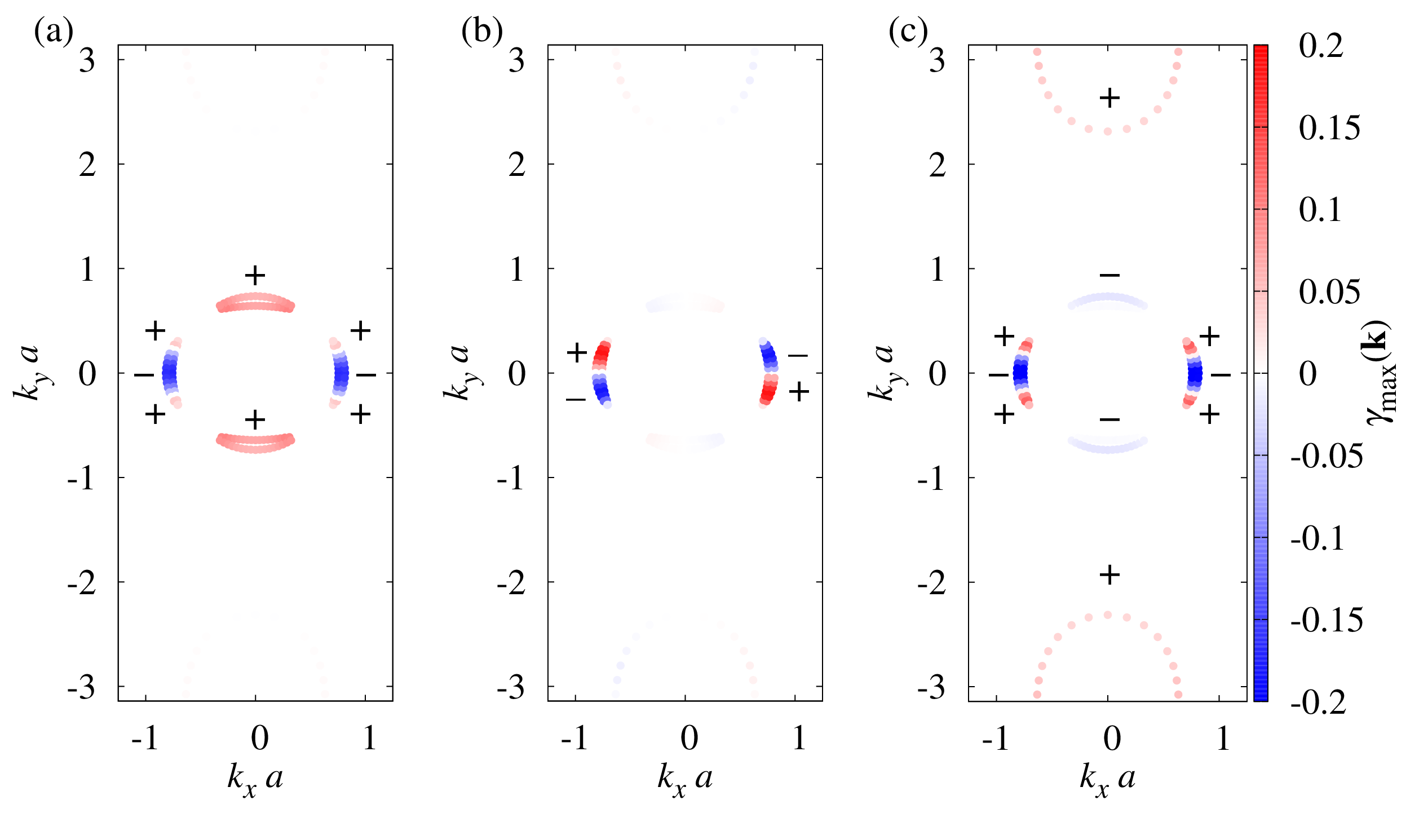}
\caption{Gap structure of (a) the dominant singlet state for
$g_{2a}/g_{cf}=0.013$, $g_1 = 0.77 t_c$, and $ g_{2b} =
0$, (b) the dominant singlet state for $g_{2a}/g_{cf}=0.022$, $g_1 = 3.49 t_c$,
and $ g_{2b} =
0$, and (c) the subdominant singlet state for $g_{2a}/g_{cf}=0.022$, $g_1 =
3.49 t_c$, and $ g_{2b} = 0$.}
\label{gap_g1-g3}
\end{figure*}

\begin{figure}
\includegraphics[clip,width=\columnwidth]{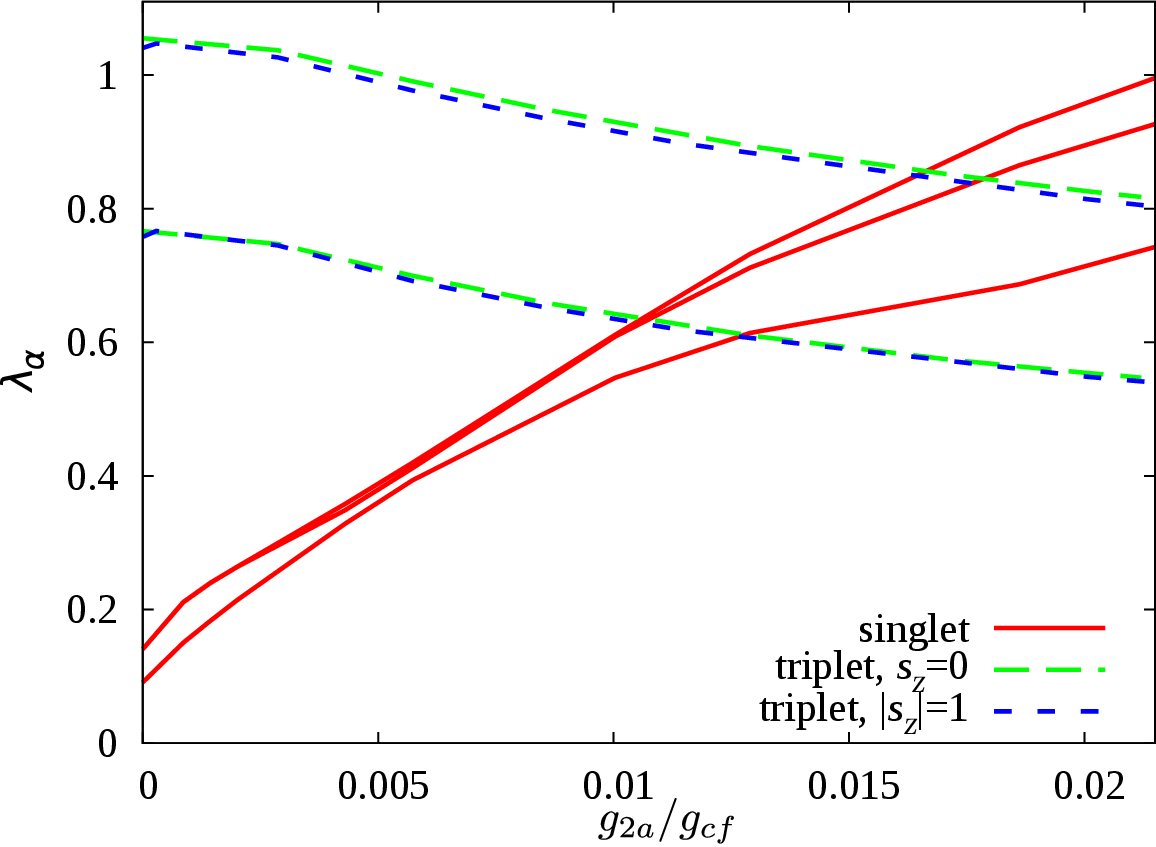}
\caption{Largest eigenvalues of the linearized gap
equation as functions of $g_{2a}/g_{cf}$.
The intraband repulsion has been set to $g_1=3.49 t_c$ and $g_{2b}=0$.}
\label{lambda_vs_g3_g1C}
\end{figure}

Figure\ \ref{lambda_vs_g3_g1C} shows the evolution of the
eigenvalues corresponding to the singlet and triplet
channels as functions of $g_{2a}/g_{cf}$ for $g_1=3.49 t_c=g_{\rm
SDW}$. It becomes clear that even when the intraband Coulomb repulsion
$g_1$ takes a
rather large value, a small $g_{2a}$ is sufficient to make a
quasi-spin-singlet state the dominant SC instability.
The leading singlet state then has the structure
shown in Fig.\ \ref{gap_g1-g3}(b). It is
closely followed by a state with the gap depicted
in Fig.\ \ref{gap_g1-g3}(c), illustrating the tendency of the
intraband repulsion to favor nodal gap structures. The close proximity
of the eigenvalues for the two different gap structures suggests that 
small changes in the model, e.g., in
the band structure, may change the order of the two eigenvalues.

\subsection{Interband-hopping transitions $g_{2b}$}

\begin{figure}[b]
\includegraphics[clip,width=\columnwidth]{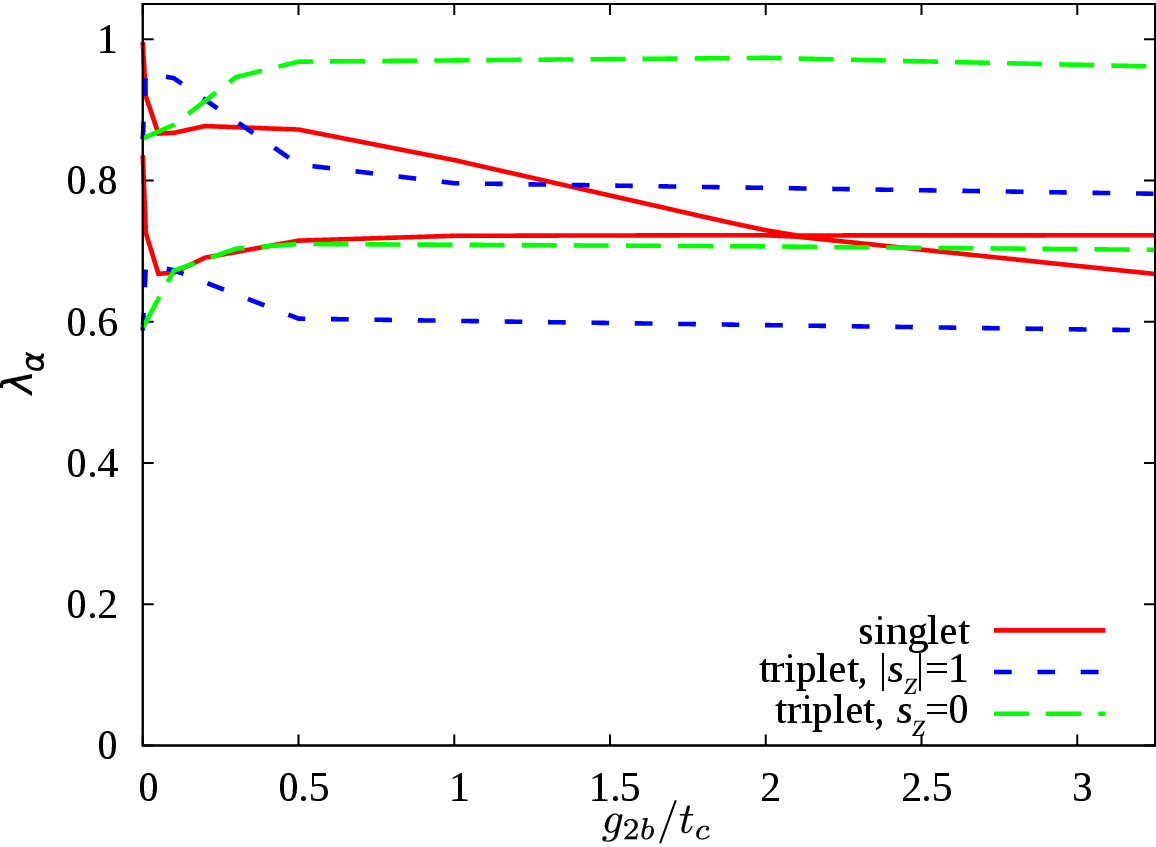}
\caption{Largest eigenvalues of the linearized gap equation in the
singlet and triplet channels, as functions of $g_{2b}$.
The ratio of the pair hopping and the interband repulsion has been set to
$g_{2a}/g_{cf}=0.013$ with $g_{\rm SDW}=3.49 t_c$ and $g_1=0$.}
\label{lambda_vs_g2b}
\end{figure}

Finally, we study the effect of the second type of correlated interband-hopping
transition with coupling constant $g_{2b}$, given in Eq.\ (\ref{g_2b}). We take
$g_{2a}/g_{cf}=0.013$ as in Sec.\ \ref{intraband_coul}, set $g_1=0$, and vary
$g_{2b}$. As with the correlated pair hopping $g_{2a}$, the SDW order 
is only stable for non-negative values of
$g_{2b}$.~\cite{PhysRevB.24.5713,PhysRevB.80.174401} Furthermore, it has been
pointed
out that for the iron pnictides the inequality
$g_{2b}<g_{cf}$ is most likely satisfied.\cite{PhysRevB.78.134512}
Therefore, we only consider the  interval $0 \le g_{2b} < g_{cf} = 3.445 t_c$.

In Fig.\ \ref{lambda_vs_g2b} we plot the largest eigenvalues from the SC gap
equation in the singlet and triplet channels as functions of $g_{2b}$.
This shows that a finite $g_{2b}$ breaks the near degeneracy of the
triplet states with $s_z=0$ and $|s_z|=1$. We also find that a non-zero $g_{2b}$
leads to a strong
suppression of the $s^\pm$-type state: Almost immediately upon
switching on $g_{2b}$, the $p_x$-wave state
becomes the dominant SC state again. Similarly to the effect of the
intraband repulsion $g_1$, we see from Fig.\ \ref{IA_g1}(b) that $g_{2b}>0$
leads to a reduction of the
repulsion between the small electron and hole pockets, hence reducing
the tendency towards a sign change between these pockets and therefore
suppressing the $s^\pm$-type pairing. 
 At $g_{2b}\approx 2.1 t_c$, the two largest eigenvalues in the
singlet channel cross and a state with the structure
shown in Fig.\ \ref{gap_g1-g3}(b) becomes the dominant singlet
state again.

\section{Summary and Conclusions}\label{summary}

We have presented a method that allows us to derive an effective pairing
interaction for a multiband system in a symmetry-broken SDW phase. Our
approach is to decouple an interacting multiband system with general
two-particle interactions in the spin channel and to apply a saddle-point
approximation to describe the SDW phase. The remaining fluctuations in
the decoupling field are integrated out to obtain the quasiparticle interactions
in the ordered phase. In the presence of the SDW, we calculate the
susceptibilities for transverse and longitudinal particle-hole excitations
within the RPA. These susceptibilities determine the effective pairing
interaction in the quasi-spin-singlet and quasi-spin-triplet channels. The
pairing interactions are then inserted into the linearized gap equation in
order to find the leading SC instability. This approach allows us to study the
effect of spin and charge fluctuations on the pairing in the SDW phase.
In particular, it is an unbiased tool for finding the gap structure of the
leading SC instability since the gap structure is obtained as an eigenvector
from the gap equation. In this respect our
approach is advantageous compared to Ginzburg-Landau
calculations that only allow for a limited number of different gap structures.

We have applied this approach to a two-band minimal model for iron
pnictides. The effective pairing interaction has been calculated
for various combinations of four symmetry-allowed types of interactions:
interband and intraband Coulomb repulsion and two types of correlated
interband-hopping terms.
Our results show that there is a complex interplay between the bare
interactions, the susceptibilities, and the transformation factors that arise
from the folding of the BZ in the SDW phase. The description of this interplay
 is the key difference of our approach compared to previous microscopic
approaches to describe the coexistence region in the pnictide phase diagram. The
effect of the electron-electron interactions on the pairing is not included in a
spin-fermion model.\cite{0953-8984-23-9-094203} Decoupling the bare interaction
within a mean-field approximation\cite{PhysRevB.80.100508, PhysRevB.83.224513,
PhysRevB.81.174538} neglects the crucial role of fluctuations in promoting the
pairing.

Note that although the interband components of the transverse spin 
susceptibility diverge, the magnons do not lead to a divergence in the effective
pairing interaction. 
The fluctuation-enhanced interaction leads to the appearance of a
quasi-spin-triplet $p_x$-wave pairing state that is not found if only the bare
interactions are considered. 
The $p_x$-wave state competes with the quasi-spin singlet states,
which are much more sensitive to the strengths of the bare
interactions. In particular, a finite pair-hopping 
amplitude $g_{2a}$ is crucial for the formation of 
singlet pairs, and the singlet eigenvalues react sensitively to
changes in the ratio $g_{2a}/g_{cf}$.  We expect that this competition can be 
found also in the spin-fermion model proposed by Wu and
Phillips\cite{0953-8984-23-9-094203} because the key features of the spin
susceptibility are present in both models. However, unlike the itinerant
picture, the spin-fermion model is based on the assumption of localized spins.
Hence, the physical basis of the two models is quite different and there is no
direct mapping between the interaction strengths in our Hamiltonian and the
parameters of the spin-fermion model.

Although $g_1$ and $g_{2b}$
suppress the singlet pairing, the parameter range in which a
triplet state is the dominant instability is limited,
as a small increase in $g_{2a}/g_{cf}$ always leads to a dominant singlet
state. For $g_1=g_{2b}=0$ and $g_{2a}/g_{cf}\gtrsim 0.005$, the singlet channel
is clearly
dominated by a nodeless $s^\pm$-type state suggested to be the most
likely pairing state in earlier works.\cite{PhysRevB.81.174538,
PhysRevB.82.014521, PhysRevB.84.180510} However, if either 
$g_1$ or $g_{2b}$ are sufficiently large, nodal gap
structures are favored. The dominant state for large $g_1$ or $g_{2b}$ has 
nodes along the $k_x$ and $k_y$ axes.

In conclusion, we find that a nodeless $s^\pm$-type singlet pairing
state, several nodal singlet states, and a $p_x$-wave triplet state can be the
leading SC instability in the SDW phase of a two-band model for the
iron pnictides. The dominant instability depends sensitively on the four
coupling strengths. Hence, these coupling strengths could be constrained by the
experimental determination of the gap structure in the coexistence
region, which has hitherto not been studied in much detail. Although
there are reports of a transition from a nodal to a nodeless state in
$\mathrm{Ba}_{1-x}\mathrm{K}_x\mathrm{Fe_2As_2}$ with
decreasing hole doping based on thermal-conductivity
measurements,\cite{arXiv:1105.2232} it is unclear where these nodes appear
on the Fermi surface. Momentum-resolved measurements of the gap to
distingusih between the different structures are
therefore highly desirable, with angle-resolved
photoemission spectroscopy (ARPES) being
the method of choice. The transition from
a nodal to a nodeless structure was explained by Maiti \textit{et
al.}\cite{PhysRevB.85.144527} as a result of the change in the SDW gap size
with doping. Our work suggests an alternative
explanation: we find that the gap structure depends strongly on details of the
interactions. In view of our results it is intriguing that the nodes appear
when the hole concentration is reduced.\cite{arXiv:1105.2232} A
reduction of the hole concentration is expected to increase the
effective Coulomb repulsion in our Hubbard-type model due to
weaker screening. This should result in an increase of
the intraband Coulomb repulsion $g_1$ relative to the SDW interaction $g_{\rm
SDW}$, which we find to stabilize nodal singlet states.

\section*{Acknowledgments}

The authors thank M. Breitkreiz, M. Daghofer, A. F. Kemper, D. K. Morr, S.
Sachdev, and A. Vishwanath for useful discussions. J. S. is grateful for the
hospitality of the University of California, Berkeley, where this work was
initiated. Financial support by the Deutsche Forschungsgemeinschaft through
Priority
Programme SPP 1458 and Research Training Group GRK 1621 is acknowledged.

\end{document}